\documentclass[a4paper,11pt]{article}
\pdfoutput=1 

\usepackage{jheppub} 
\usepackage[compat=1.1.0]{tikz-feynman}
\usepackage{graphicx, subcaption}
\usepackage[T1]{fontenc} 
\usepackage{comment}
\usepackage{multirow}

\DeclareRobustCommand{\Fig}[1]{Fig.~\ref{#1}}

\title{Non-resonant Anomaly Detection \\
with Background Extrapolation}

\author[1,2]{Kehang Bai,}
\author[2,3]{Radha Mastandrea,}
\author[3,4]{and Benjamin Nachman}

\affiliation[1]{Institute for Fundamental Science and Department of Physics, \\ University of Oregon, Eugene, OR 97403, USA}
\affiliation[2]{Physics Division, Lawrence Berkeley National Laboratory, Berkeley, CA 94720, USA}
\affiliation[3]{Department of Physics, University of California, Berkeley, CA 94720, USA}
\affiliation[4]{Berkeley Institute for Data Science, University of California, Berkeley, CA 94720, USA}

\emailAdd{kbai@uoregon.edu}
\emailAdd{rmastand@berkeley.edu}
\emailAdd{bpnachman@lbl.gov}

\abstract{Complete anomaly detection strategies that are both signal sensitive and compatible with background estimation have largely focused on resonant signals.  Non-resonant new physics scenarios are relatively under-explored and may arise from off-shell effects or final states with significant missing energy.  In this paper, we extend a class of weakly supervised anomaly detection strategies developed for resonant physics to the non-resonant case.  Machine learning models are trained to reweight, generate, or morph the background, extrapolated from a control region.  A classifier is then trained in a signal region to distinguish the estimated background from the data.  The new methods are demonstrated using a semi-visible jet signature as a benchmark signal model, and are shown to automatically identify the anomalous events without specifying the signal ahead of time.}

\begin{document} 
\maketitle
\flushbottom

\section{Introduction}
\label{sec:intro}

Despite the impressive predictability of the Standard Model (SM), it is a well-known fact that it does not account for all known phenomena and is plagued with a number of aesthetic issues.  As such, the search for new, fundamental interactions is a central goal of particle physics across virtually every experiment.  However, given that there is an uncountable number of new physics models, we simply do not have the bandwidth to test all possibilities, and it is certain that there are alternative hypotheses that we have not yet considered.

For this reason, a new paradigm has emerged to complement traditional, model-specific approaches.  The new \textit{anomaly detection} protocols seek to explore data with as little bias as possible in order to be broadly sensitive to many scenarios. Such protocols have been significantly advanced by modern machine learning (see~\cite{Kasieczka:2021xcg,Aarrestad:2021oeb,Karagiorgi:2021ngt,Feickert:2021ajf} for a selection of machine learning reviews and anomaly detection challenges, and~\cite{Collins:2018epr,DAgnolo:2018cun,Collins:2019jip,DAgnolo:2019vbw,Farina:2018fyg,Heimel:2018mkt,Roy:2019jae,Cerri:2018anq,Blance:2019ibf,Hajer:2018kqm,DeSimone:2018efk,Mullin:2019mmh,1809.02977,Dillon:2019cqt,Andreassen:2020nkr,Nachman:2020lpy,Aguilar-Saavedra:2017rzt,Romao:2019dvs,Romao:2020ojy,knapp2020adversarially,collaboration2020dijet,1797846,1800445,Amram:2020ykb,Cheng:2020dal,Khosa:2020qrz,Thaprasop:2020mzp,Alexander:2020mbx,aguilarsaavedra2020mass,1815227,pol2020anomaly,Mikuni:2020qds,vanBeekveld:2020txa,Park:2020pak,Faroughy:2020gas,Stein:2020rou,Chakravarti:2021svb,Batson:2021agz,Blance:2021gcs,Bortolato:2021zic,Collins:2021nxn,Dillon:2021nxw,Finke:2021sdf,Shih:2021kbt,Atkinson:2021nlt,Kahn:2021drv,Dorigo:2021iyy,Caron:2021wmq,Govorkova:2021hqu,Kasieczka:2021tew,Volkovich:2021txe,Govorkova:2021utb,Hallin:2021wme,Ostdiek:2021bem,Fraser:2021lxm,Jawahar:2021vyu,Herrero-Garcia:2021goa,Aguilar-Saavedra:2021utu,Tombs:2021wae,Lester:2021aks,Mikuni:2021nwn,Chekanov:2021pus,dAgnolo:2021aun,Canelli:2021aps,Ngairangbam:2021yma,Bradshaw:2022qev,Aguilar-Saavedra:2022ejy,Buss:2022lxw,Alvi:2022fkk,Dillon:2022tmm,Birman:2022xzu,Raine:2022hht,Letizia:2022xbe,Fanelli:2022xwl,Finke:2022lsu,Verheyen:2022tov,Dillon:2022mkq,Caron:2022wrw,Park:2022zov,Kamenik:2022qxs,Hallin:2022eoq,Kasieczka:2022naq,Araz:2022zxk,Mastandrea:2022vas,Schuhmacher:2023pro,Roche:2023int,Golling:2023juz,Sengupta:2023xqy,Mikuni:2023tok,Golling:2023yjq,Vaslin:2023lig,ATLAS:2023azi,Chekanov:2023uot,CMSECAL:2023fvz,Bickendorf:2023nej,Finke:2023ltw,Buhmann:2023acn,Freytsis:2023cjr} for specific techniques).  One powerful anomaly detection strategy (``weak supervision'') is to isolate a region of phase space and compare data to a background-only reference.  Signal model-agnostic evidence for new physics emerges when the data is not statistically consistent with the reference.

The core task of weakly supervised anomaly detection strategies is to construct the reference sample of background-only events.  Existing strategies follow one of two approaches: (1) the reference sample is from (background-only) simulation and (2) the reference sample is learned from sidebands.  The advantage of the simulation approach is that it puts few assumptions on the form of the signal. In particular, this approach can accommodate signals that are non-resonant, either due to the new physics being very massive or consisting of non-reconstructable (e.g. invisible) particles.  The challenge with a simulation-based reference is that the simulation accuracy limits the anomaly detection sensitivity.  In contrast, sideband methods posit that the signal, if it exists, is localized in at least one known dimension $m$.  Focusing on a particular interval in $m$, a signal-sensitive region of phase space is then constructed using other features $x$, and regions in $m$ away from the posited resonance are used to estimate the background $p_\text{background}(x|m)$.  The benefit of sideband methods is that the reference is estimated directly from data, while the challenge is that not all signals are resonant.

We propose an approach to merge the advantages of the simulation-based and sideband-based cases to form a strategy that learns the reference from data without requiring a resonance.  This method extends and modifies existing resonance approaches~\cite{Andreassen:2020nkr,Nachman:2020lpy,Hallin:2021wme,Golling:2022nkl,Hallin:2022eoq,1815227,Sengupta:2023xqy} to allow for extrapolation in $m$, as opposed to interpolation.  Instead of the signal region being defined by an interval in a one-dimensional $m$, we now have a multidimensional $m$ for which the signal region is defined by thresholds\footnote{A non-rectangular signal region is also possible, but makes the accounting more difficult.} $m_0>c_0,m_1>c_1,...,m_n>c_n$ for $m,c\in\mathbb{R}^n$.  The conditional reference $p_\text{background}(x|m)$ is learned (directly or indirectly) from the complement of the signal region and then extrapolated to the signal region.  If the machine learning models are smooth and the background remains qualitatively the same in the signal region (e.g. the functional form of the background distribution, particularly the dependence on the context variables,  is continuous across the CR-SR boundary), there is reason to believe that the extrapolation can be accurate\footnote{We leave additional studies to impose minimalism in the extrapolation to future work via e.g. monotonicity~\cite{Kitouni:2021fkh}.}.  A new feature of the non-resonant case is that we need to additionally estimate $p_\text{background}(m)$.  This is also needed in the resonant case, but it is less important (since $p(m)$ is relatively constant in the signal region) and thus approximate or simplified methods have been considered so far.  We propose to estimate $p_\text{background}(x|m)$ using a likelihood-ratio method~\cite{Andreassen:2020nkr} and then combine this estimate with three proposals for estimating $p_\text{background}(m)$ based on reweighting~\cite{Andreassen:2020nkr}, generative models~\cite{Hallin:2021wme}, and template morphing~\cite{Golling:2022nkl}.

The growing anomaly detection for fundamental interactions literature includes a number of proposals that can accommodate non-resonant new physics.  In the weakly supervised case, simulation-based background estimates do not require resonant new physics~\cite{DAgnolo:2018cun,DAgnolo:2019vbw,dAgnolo:2021aun,Letizia:2022xbe}.  Nearly all data-based weakly supervised methods include non-resonant $x$ features~\cite{Andreassen:2020nkr,Nachman:2020lpy,Hallin:2021wme,Golling:2022nkl,Hallin:2022eoq,1815227,Sengupta:2023xqy,Bickendorf:2023nej}.  In special cases, symmetries can be used to define the reference directly in data without requiring $m$ to be resonant~\cite{Tombs:2021wae,Lester:2021aks,Birman:2022xzu,Finke:2022lsu}. Extrapolation of generative models has been considered for background estimation, but combined with classical model-specific search strategies~\cite{Lin:2019htn}.  Unsupervised methods like those based on autoencoders (e.g. Ref.~\cite{Hajer:2018kqm,Farina:2018fyg,Heimel:2018mkt} and many others) do not require resonances for achieving signal sensitivity in general. However, except in special cases, they often lack background estimation strategy~\cite{Mikuni:2021nwn}, or any guarantees of optimality. Our approach is unique because it combines signal sensitivity with background estimation\footnote{The reference does not need to be used directly to estimate the background after making a cut on the anomaly score.  One could use any classical non-resonant background estimation strategy like the ABCD method, possibly also with machine learning~\cite{Kasieczka:2020pil}.} and is asymptotically optimal~\cite{Nachman:2020lpy} in the limit of large datasets and effective machine learning.

Non-resonant signals can arise from many new physics models. For example, models of a strongly interacting dark sector can produce jets which contain both stable and unstable dark hadrons, predicting signatures of jets containing significant missing energy. Non-resonant signals are also characteristic of effective field theories, which can model new physics with masses beyond those accessible through on-shell production. ML-based anomaly detection can offer additional sensitivity to explore the vast signature space of non-resonant signals.

This paper is organized as follows. In Sec.~\ref{sec:methodology}, we introduce the methodology we follow for our tail-based non-resonant anomaly detection procedure, providing details on the background extrapolation task.
In Sec.~\ref{sec:physics_results}, we show a realistic physics example, considering jets emerging from a complex dark sector as our non-resonant signal. We conclude in Sec.~\ref{sec:conclusions}.

\section{Methodology}
\label{sec:methodology}

\subsection{Locating overdensities}

The goal of a weakly-supervised anomaly search is to identify regions of phase space that are more represented by data (forming overdensities) than they are by the reference dataset.  The search is performed in a signal region (SR) and the reference is estimated using information from a neighboring region. 

In the resonant case, the SR is defined by an interval around a posited new particle mass $m\in [m_0-\delta,m_0+\delta]$.  The neighboring region (called \textit{sideband} or SB) is the region (or a subset of the region) outside of this interval.  If the signal exists and is at mass $m_0$, we expect that most of the signal is in the SR with little contamination in the SB.  The reference background template is estimated using information from the SB.  Since we expect the phase space to be smoothly varying with $m$ and since $\delta$ is expected to be small relative to regions over which large changes occur in $p_\text{background}(x|m)$, we expect the background to be well-estimated by interpolating from the SB (see further details in Sec.~\ref{sec:background}).  Given samples from the background dataset\footnote{There are variations on this where the density is used directly without samples~\cite{Nachman:2020lpy}, but the overall idea is the same.}, overdensities are identified with the Classification WithOut LAbels procedure (\textsc{CWoLa})~\cite{Metodiev:2017vrx,Collins:2018epr,Collins:2019jip}.  Events in data are all assigned the label ``signal'' and all events from the background template are labeled ``background''.  An optimal classifier trained to distinguish these datasets will implicitly learn a function monotonically related to the signal-over-background likelihood ratio.  Thresholding the classifier will then isolate an anomaly-like region of phase space.

\begin{figure}[ht]
    \centering
    \includegraphics[width = .6\textwidth]{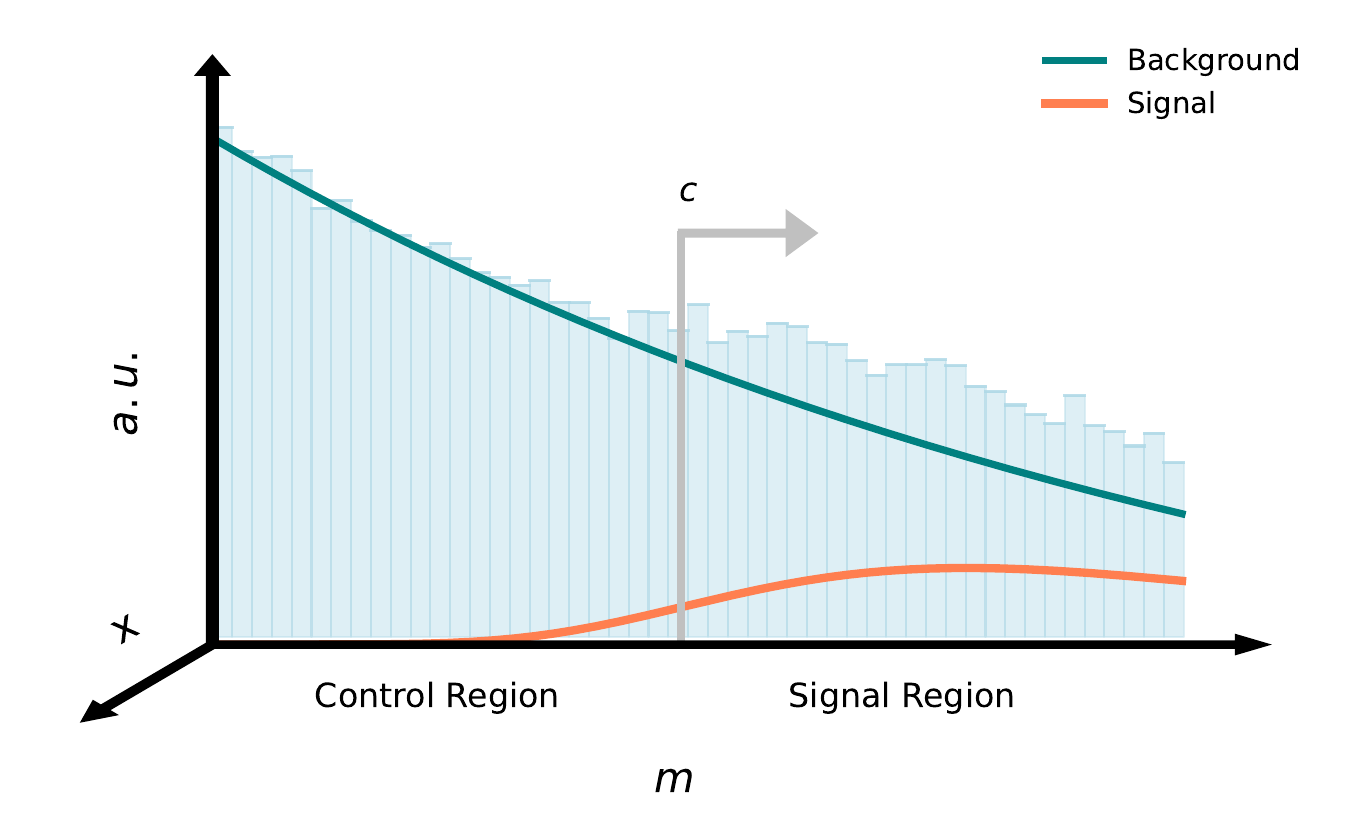}
    \caption{A schematic of the setup for non-resonant anomaly detection. The signal region (SR) is defined by a one-sided cut on a context variable $m > c$. While the figure shows a one-dimensional context variable, in practice the context can be multidimensional. A number of other features $x$ are used for the \textsc{CWoLa} anomaly detection classifier.  It is also possible to include $m$ in the classifier.}
    \label{fig:non-resonant_schematic}
\end{figure}

In the non-resonant case, the setup is slightly different.  Instead of having SB that surround the SR, the neighboring region is only on one side and is now called a \textit{control region} or CR.  However, we allow for $m$ to be multidimensional -- the more dimensions available to constrain $p_\text{background}(x|m)$, the better accuracy we might expect for reconstruction in the SR\footnote{We validated this with simplified Gaussian samples.}.  Resonances are in fact nearly a special case of this setup, since one could define e.g. $m' = 1/|m-m_0|$.  As in the resonant case, we start by constructing background templates in the CR, and we expect most of the signal to be in the SR so that the CR can be used to estimate $p_\text{background}$ without bias. A schematic diagram of the non-resonant setup is presented in  \Fig{fig:non-resonant_schematic}.  The challenge now is to extrapolate into the SR instead of interpolating as in the resonant case.  Given the background template, the actual anomaly detection proceeds using the same \textsc{CWoLa} procedure described above.  The main aspect that distinguishes different approaches is how the background templates are generated.

\

\subsection{Constructing background templates}
\label{sec:background}

To construct samples of background-only events in the SR, we take the following steps:

\begin{enumerate}
    \item \textbf{Estimate $p_{\mathrm{data}}(x|m)$ in the CR.} We first learn the distribution of the chosen features for background-only events, conditioned on the context variables, in the CR. It is assumed that the conditioning on the context will allow the learned distribution to be extrapolated into the SR. For the interpolation case (i.e resonant anomaly detection), the analogous first step would be to learn $p_{\mathrm{data}}(x|m)$ in the SB.  Since we assume that the signal is mostly absent from the CR, $p_\text{data}\approx p_\text{background}$. 
    
    \item \textbf{Estimate $p_{\mathrm{background}}(m)$ in the SR.} In order to sample from $p_{\mathrm{background}}(x|m)$ in the SR, we need to be able to generate SR-like context $m$. In the interpolation case, this might be done by performing a parametric fit to a histogram of $m$ in the SB and interpolating to the SR.  This could also work in the extrapolation case, but (1) the shape may be non-trivial in the tails of $m$ and (2) this approach does not scale well to more than one dimension.  
    
    To circumvent these challenges, we estimate $p_{\mathrm{background}}(m)$ in the SR by reweighting context in the CR, $p_{\mathrm{simulation}}(m)$, to match data and then extrapolating this function into the SR.  A binary classifier parameterized in $m$~\cite{Cranmer:2015bka,Baldi:2016fzo} is trained to distinguish simulated context from data context in the CR.  The classifier output is then interpreted as a likelihood ratio $w(m)$ for estimating the background~\cite{Andreassen:2020nkr}. As in step 1, we assume that conditioning the network on the context will enable the learned weights to extrapolate into the SR.

    \item \textbf{Sample $p_\mathrm{background}(x|m)$ in the SR}. We first estimate $p_{\mathrm{background}}(x|m)$ from the learned model in the CR, drawing context from $p_{\mathrm{simulation}}(m)$, in the CR. The resulting events are then weighted by $w(m)$ to arrive at our estimate of $p_\text{background}(x,m)$ in the SR, following Eq.~\ref{eq:reweighting}.
    
    \end{enumerate}

\begin{equation}
\label{eq:reweighting}
\begin{split}
    p_{\rm background}(x,m) 
    &= p_{\rm background}(x|m)p_{\rm background}(m) \\
    &= p_{\rm background}(x|m)p_{\rm simulation}(m)\frac{p_{\rm background}(m)}{p_{\rm simulation}(m)} \\ 
    &= p_{\rm background}(x|m)p_{\rm simulation}(m)w(m)
\end{split}
\end{equation}

\noindent We consider three approaches for approximating $p_{\mathrm{data}}(x|m)$ in the CR, based on three techniques previously developed for the resonance case: Simulation-Assisted Likelihood-Free Anomaly Detection (SALAD) \cite{Andreassen:2020nkr}, Classifying Anomalies through Outer Density Estimation (CATHODE) \cite{Hallin:2021wme}, and Flow-Enhanced Transportation for Anomaly detection (FETA) \cite{Golling:2022nkl}. Our proposed methods are:

\begin{enumerate}
    \item \textbf{Reweight} (SALAD-inspired). The parameterized, binary classifier used to approximate $p_\text{background}(m)$ is extended to include also $x$ so that the likelihood ratio in both $x$ and $m$ is estimated at the same time. 
    \item \textbf{Generate} (CATHODE-inspired). A generative neural network (normalizing flow~\cite{pmlr-v37-rezende15}) is trained to model data in the CR, conditioned on the context $m$. Given values of $m$, one can sample features $x$ from the normalizing flow.
    \item \textbf{Morph} (FETA-inspired). A normalizing flow-based model is trained to map samples in the CR from MC to data, conditioned on the context $m$.  Like SALAD, FETA starts from a simulation, but instead of reweighting, it morphs the features directly~\cite{Golling:2023mqx,Bright-Thonney:2023sqf}.
    
\end{enumerate}
Note that $w(m)$ is estimated with SALAD in all three cases, so our new approaches are SALAD-\{SALAD, CATHODE, FETA\} hybrid methods. 

We evaluate the performance of the Reweight, Generate, and Morph-constructed background samples against two benchmarks: (1) a fully supervised classifier trained to discriminate pure background from pure signal, and (2) an ``idealized'' classifier trained to discriminate pure background from true background $+$ signal. The idealized classifier represents the actual best performance to which our background estimation methods should asymptote.

\subsection{Network architectures and training specifications}
\label{sec:NN_specifications}

All networks are implemented in \textsc{PyTorch} \cite{NEURIPS2019_9015} and optimized with \textsc{Adam} \cite{kingma2017adam}. For the Generate and Morph methods, we construct normalizing flow networks with the \textsc{nflows} package~\cite{nflows}. All models are trained with a train-validation split of $2/3$--$1/3$ and are evaluated at the epoch of lowest validation loss. For the Reweighting, Generate, and Morph methods, architectures and hyperparameters were optimized through manual tuning such that in the case of zero signal injection, a binary classifier could not distinguish reconstructed background in the CR from data in the CR. This was necessary in order to keep the SR blinded.

The Reweighting method is implemented with a binary classifier, consisting of a dense neural network with 3 layers of 100 nodes each. We use a batch size of 512, a learning rate of $10^{-3}$, and train for up to 50 epochs with an early stopping patience of 5. The classifier that generates the context weights $w(m)$ is exactly the same as that for the Reweighting method. The Generate method is implemented with 4 layers of (Masked Affine Autoregressive Transform block of 128 hidden features, Reverse Permutation), where the transform block is built from a Masked Autoencoder for Distribution Estimation (MADE) architecture. Training is with a batch size of 216, a learning rate of $10^{-3}$, and is carried out for up to 20 epochs with an early stopping patience of 5 epochs. For the Morph method, the base density flow uses the same architecture as the Generate method. The transfer flow is similar, but layers have only $64$ hidden features, the batch size is reduced to 128, and the learning rate to $10^{-4}$. 

For the anomaly detection classifier networks, as well as the networks used for the closure tests (see Sec. \ref{sec:phys_closure_tests}), we use a fully-connected architecture with 3 layers of 64 nodes each. We use a batch size of 512, a learning rate of $10^{-3}$, and train for up to 50 epochs with an early stopping patience of 5 epochs. Networks are only trained on the non-context features. In order to ensure training stability, for a given set of generated samples, we crop the weights $w(m)$ by dropping all weights that are greater than 3 standard deviations above the mean. This has the effect of removing a very small number (usually on the order of a few tens, or $\sim 0.1\%$) of samples whose weights severely bias the resulting $p_{\rm background}(x,m)$ distributions.

For the dataset, we consider a two-dimensional context $m$ and a five-dimensional feature space $x$ (to be explained in detail in Sec. \ref{sec:phys_data_spec}). The values of all seven variables are minmax-scaled to the range $(-2.5, 2.5)$ based on the minimum and maximum of the simulation (pure background) dataset before any network training is carried out. This was found to give more reliable flow training. When the number of signal events is small, the exact events that are injected into the ``data'' is important.  Therefore, we rerun the procedures with $10$ random signal injections and report the median and a $68$-percentile spread.  For each signal injection, we retrain the models with a different random initialization $20$ times and average over these models.  We found (as in previous studies) that this ensembling helps with sensitivity and stability.

\section{Application with dark QCD jets}
\label{sec:physics_results}

This section illustrates the use of the above background extrapolation method in the case of finding Beyond the Standard Model (BSM) physics at the LHC. The physics model we are interested in is jets that contain both stable and unstable hadrons from a strongly interacting dark sector, or dark QCD. This type of signal comes from e.g. Hidden Valley models~\cite{Strassler:2006im,Carloni:2010tw,Carloni:2011kk,Knapen:2021eip}. In particular, we assume there exist some new massive quarks charged under dark QCD that are promptly produced and go through QCD-like showering and hadronization to form ``dark hadrons''. It is possible that a fraction of these dark hadrons decay back to the SM hadrons, and the rest remain invisible. This results in the type of dark QCD signature called semi-visible jets~\cite{Cohen_2015,Cohen_2017,Beauchesne_2018,Bernreuther_2021,CMS2022darkQCDsearch,atlascollaboration2023search}. Semi-visible jets are challenging to distinguish from SM QCD jets, but the unique signature from certain dark hadron mass scales and missing energy may be used for identification.

We assume that the dark quarks are produced through a mediator such as a massive gauge boson $Z'$. When the mediator is produced on-shell, we can reconstruct the invariant mass of the mediator from the semi-visible jets. However, when the invisible fraction of the jet is large, there will be a large amount of missing energy, resulting in a poor invariant mass resolution so that the reconstructed signal is no longer resonant. The invisible fraction of dark hadron decays is defined using the parameter, $r_{\rm inv}$, the ratio of the number of stable dark hadrons over number of total dark hadrons.

For this study, we choose a 4 TeV $Z'$ decay into a pair of dark quarks. There are 3 flavours of dark quarks, all with the same mass of $m_{q_{\rm D}} = 250$ GeV. There are two types of dark hadrons, the dark pion and dark rho meson, whose masses are set to be the same as the confinement scale of the dark sector, $m_{\pi_{\rm D}} = m_{\rho_{\rm D}} = \Lambda_\text{D} = 500$ GeV. Note that this choice of $ \Lambda_\text{D}$ results in a small number of dark mesons in the shower, and since the dark mesons are relatively heavy, the jet substructure comes from the dark meson decays and is less dependent on the hadronization process in the dark sector. A study of the effects from different mass parameters is presented in Appendix~\ref{appendix:a}. The AD procedure is insensitive to the simulation modeling, and therefore is complementary to methods that minimizes the hadronization uncertainties~\cite{Cohen_2020,Cohen_2023}.

\subsection{Dataset specifications}
\label{sec:phys_data_spec}

\begin{figure}[ht]
    \centering
    \includegraphics[width = 0.4\textwidth]{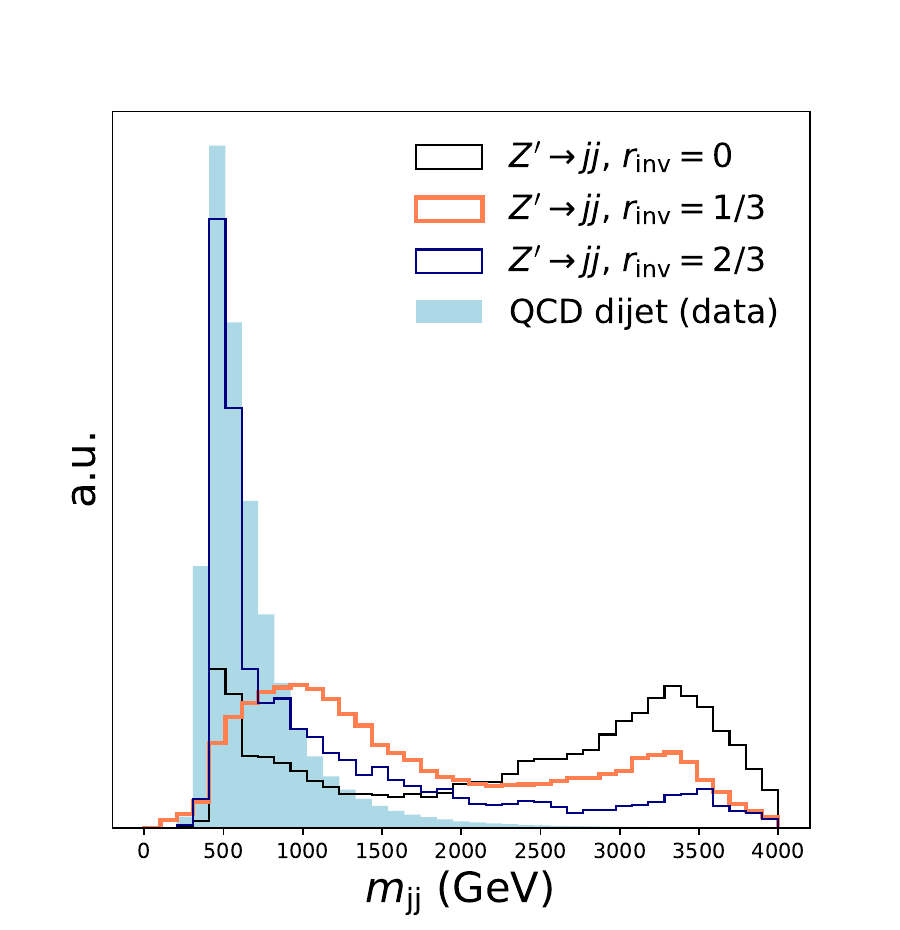}
    \caption{A histogram of $m_{jj}$, corresponding to the reconstructed mass of the $4$ TeV $Z'$ particle, for a selection of choices within the Hidden Valley model with parameter $r_{\rm inv}$. In this study, we set $r_{\rm inv} = 1/3$, at which point the $Z'$ signal is no longer clearly resonant. }
    \label{fig:mjj_scan}
\end{figure}

Signal events are generated using the Hidden Valley module from \textsc{Pythia} 8.310  \cite{bierlich2022comprehensive}. Background (``data'' and ``simulation'') events are generated using MadGraph5 aMC@NLO \cite{Alwall_2014} and showered using \textsc{Pythia} 8.310. All events use \textsc{Delphes} \cite{deFavereau:2013fsa} for detector simulation with the CMS card, and jets are clustered using the anti-$k_t$~\cite{Cacciari:2008gp} algorithm with $R = 1.0$ using \textsc{FastJet} \cite{Cacciari:2011ma}. For the background samples, the data and simulation events differ in the choices of renormalization and factorization scales (1 vs. 2, respectively) as well as the tuning (14 vs. 25). We show results using a different background simulation in Appendix~\ref{appendix:b} to validate the robustness of the anomaly detection (AD) methods against larger discrepancies in data and simulation. A minimum of 200 GeV transverse momentum is required for the leading jet in the event. In \Fig{fig:mjj_scan}, we show the $m_{\rm jj}$ distribution for a few selections of $r_{\rm inv}$ values. To ensure the non-resonance of the signal, we set $r_{\rm inv}=1/3$ in this study.

\begin{figure}[ht]
 \centering
  \begin{subfigure}[t]{.6\textwidth}
    \includegraphics[width=\textwidth]{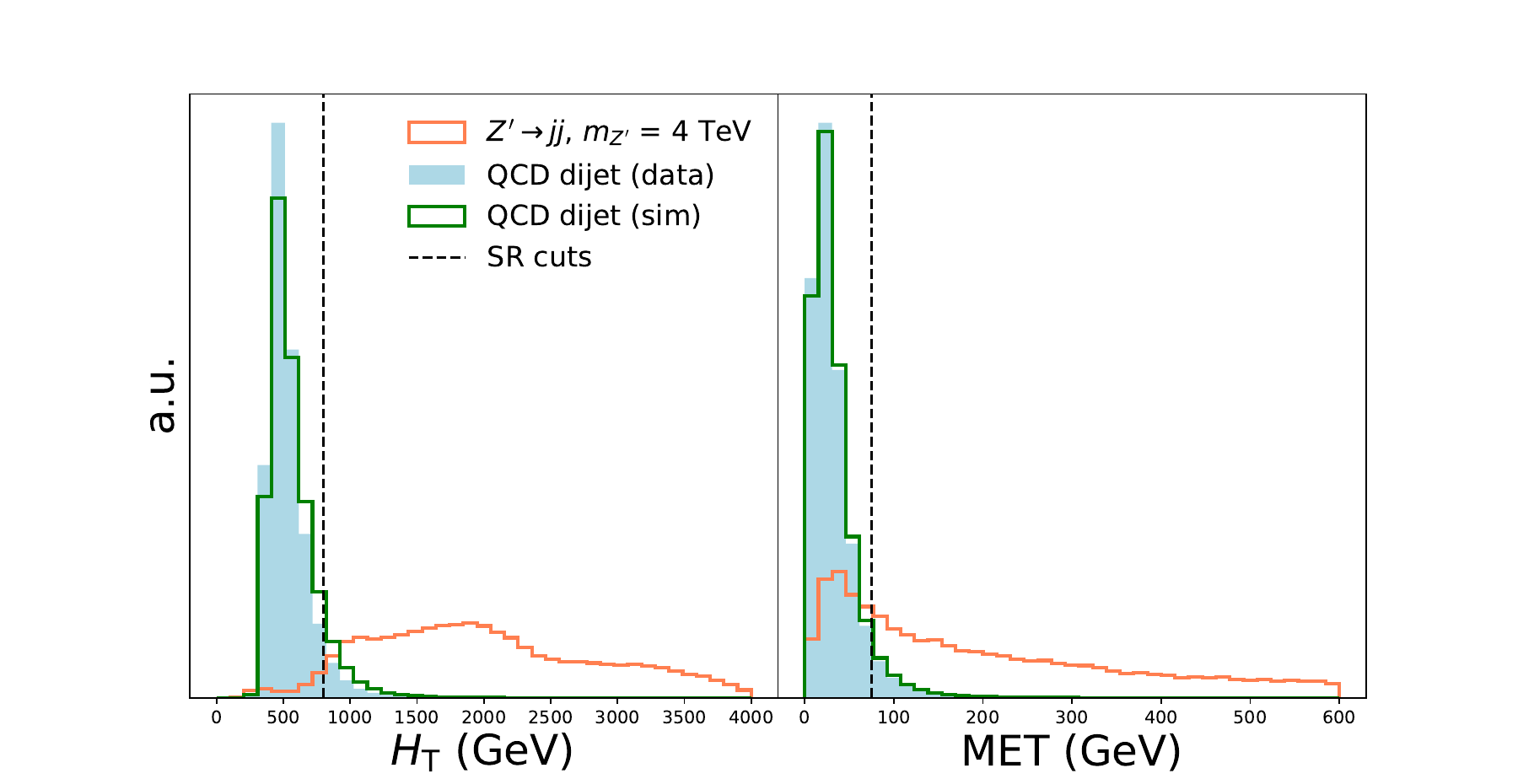}
    \caption{\label{fig:sig_vs_bkg_cont}Two-dimensional context variables. The SR is defined by the cuts $H_{\rm T} > 800$ GeV, and MET $> 75$ GeV, which are shown on the plot.}
  \end{subfigure}
  \\
  \hfill
  \begin{subfigure}[t]{1\textwidth}
    \centering
    \includegraphics[width=\textwidth]{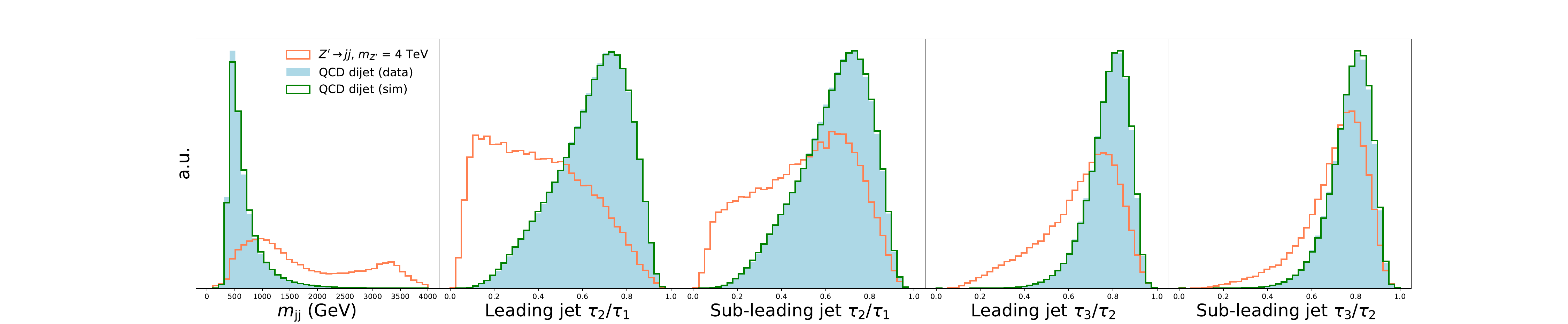}
    \caption{\label{fig:sig_vs_bkg_feat}Five-dimensional feature variables.}
  \end{subfigure}
\caption{\label{fig:physics_datasets}Histograms of the observables used in the Hidden Valley non-resonant anomaly detection task, for background (``data'' and ``simulation'') and signal events). 
}
\end{figure}

We use the scalar sum of jet transverse momenta ($H_{\rm T}$) and missing transverse momentum (MET) as the context variables to define a two-dimensional SR\footnote{These were chosen since they are nearly independent.  This is not strictly necessary for the method to work, but we found that it improves the extrapolation quality and the same features can be used for a classical matrix method / ABCD background estimate.}. For a five-dimensional feature space, we use the dijet invariant mass $m_{\rm jj}$, and the two-pronged and three-pronged N-subjetiness~\cite{Thaler:2011gf,Thaler:2010tr} for both the leading and subleading jets. Distributions of context and feature variables for signal and background is shown in \Fig{fig:physics_datasets}. The SR is defined by the cuts $H_{\rm T} > 800$ GeV, and MET $> 75$ GeV, which ensures the signal-over-background ratio is much lower in the CR than that of the SR\footnote{In addition to the selection of features, the definition of CR and SR will introduce some model dependence.  We expect that these regions will work well for a wide range of physics model parameters, especially when the $Z'$ mass is higher.}. We summarize the number of generated events in Tab.~\ref{tab:event_breakdown_physics}.

\begin{table}[ht]
    \centering
    \begin{tabular}{|c|c|c|c|}
    \hline
        Purpose & Dataset & Number CR events & Number SR events 
      \\
        \hline
        \hline
        \multirow{3}{3cm}{Training for generative models}  & Simulation (bkg.) & 9,983k & 126k  \\
        & Data (bkg.) & 9,247k & 72k \\
        & Data (sig.)  & 14k & 50k  \\
        \hline
        \multirow{ 3}{*}{Generated samples} & Reweight &  &  \\
        & Generate & N/A & 126k  \\
        & Morph &  & \\
        \hline
        \multirow{3}{*}{Evaluation} &  Ideal AD samples & N/A & 68k  \\
        & Fully supervised set & N/A & 13k bkg., 30k sig. \\
        & Test set & N/A & 150k bkg., 20k sig. \\
        \hline
        
    \end{tabular}
    \caption{Breakdown of events generated for the Hidden Valley model. We found that large training sets in the CR were necessary for good performances of the Generate and Morph (the flow-based) methods, while the Reweight method worked well even with a smaller set. We did not explore oversampling in the SR, which may improve performance~\cite{Hallin:2021wme}.}
    \label{tab:event_breakdown_physics}
\end{table}

\subsection{Results}

\subsubsection{Closure tests}
\label{sec:phys_closure_tests}

\begin{figure}
 \centering
  \begin{subfigure}[t]{\textwidth}
    \includegraphics[width=\linewidth]{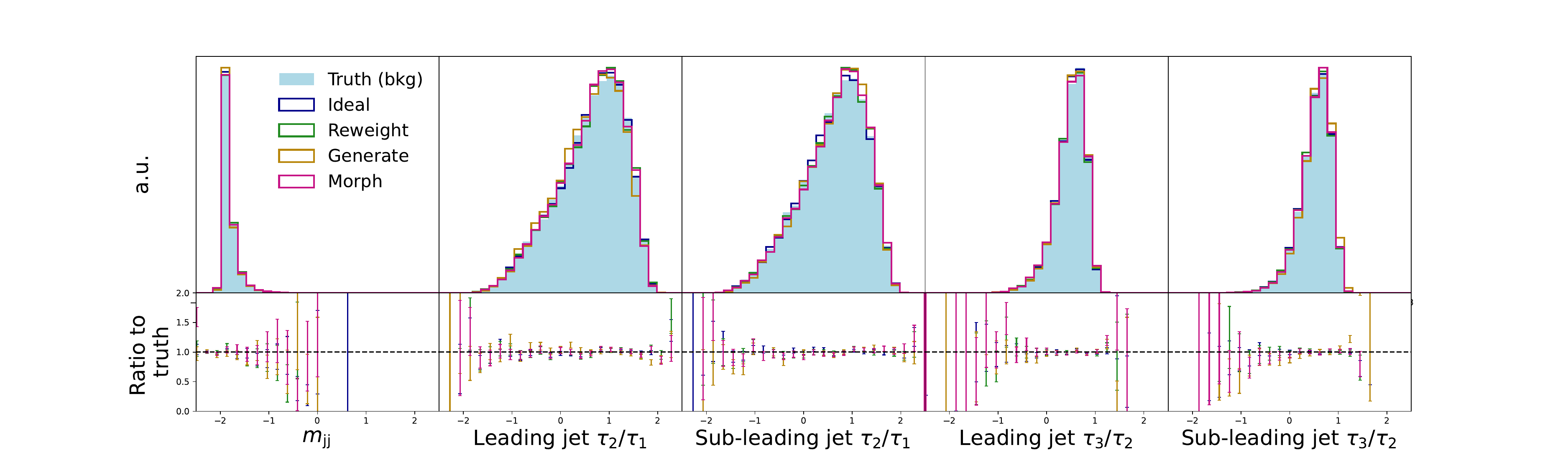}
    \caption{\label{fig:physics_features_ratios_CR}Control region distributions.}
  \end{subfigure}
  \\
  \hfill
  \begin{subfigure}[t]{\textwidth}
    \centering
    \includegraphics[width=\linewidth]{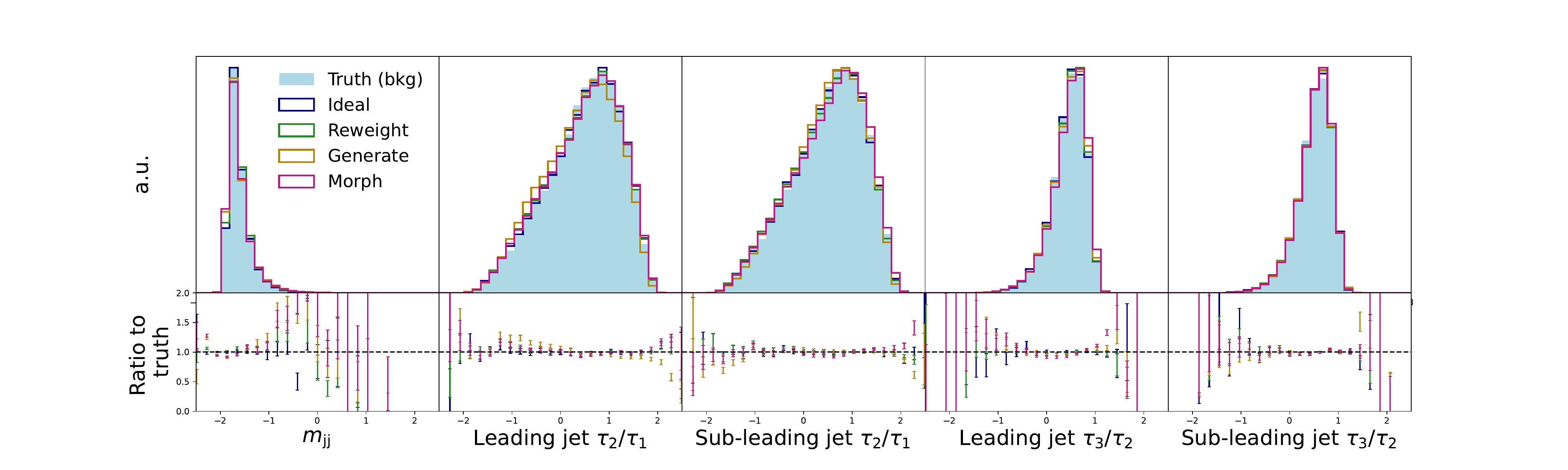}
    \caption{\label{fig:physics_features_ratios_SR}Signal region distributions.}
  \end{subfigure}
\caption{\label{fig:physics_ratios_closure} Feature distributions as constructed by the Reweight, Generate, and Morph methods in the absence of signal. Errorbars in the ratio plots reflect Poisson statistics. Note that the values of the context and features have all been minmax-scaled to the range (-2.5, 2.5).
}
\end{figure}

As a first test of extrapolation, we check that we are able to generate realistic background samples in both the CR and SR under zero-signal injection. In \Fig{fig:physics_ratios_closure}, we show the five feature distributions as constructed by the Ideal, Reweight, Generate, and Morph methods in relation to truth, the zero-signal data. Feature reconstruction is excellent in the CR, and the ratios of the marginals to the truth are consistent with unity across virtually all of the domains of the observables. Reconstruction is good overall in the SR, although the reconstructed distributions (particularly for the Morph method) deviate visibly. We calculated the one-dimensional Wasserstein distances as a quantitative measurement of the closure test, characterizing how well the observables are reconstructed in the signal region and control region by Reweight, Generate and Morph methods. The average Wasserstein distances of Ideal, Reweight, Generate, and Morph across five feature variables are $0.0048$, $0.0073$, $0.0237$, $0.0116$ in CR, and $0.0058$, $0.0130$, $0.0293$, $0.0202$ in SR, respectively. Note that in order to avoid premature unblinding, we withhold a set of 10k simulation and 10k data events in the CR from network training and use this set to evaluate CR closure.

\begin{figure}
 \centering
  \begin{subfigure}[t]{0.495\textwidth}
    \includegraphics[width=\linewidth]{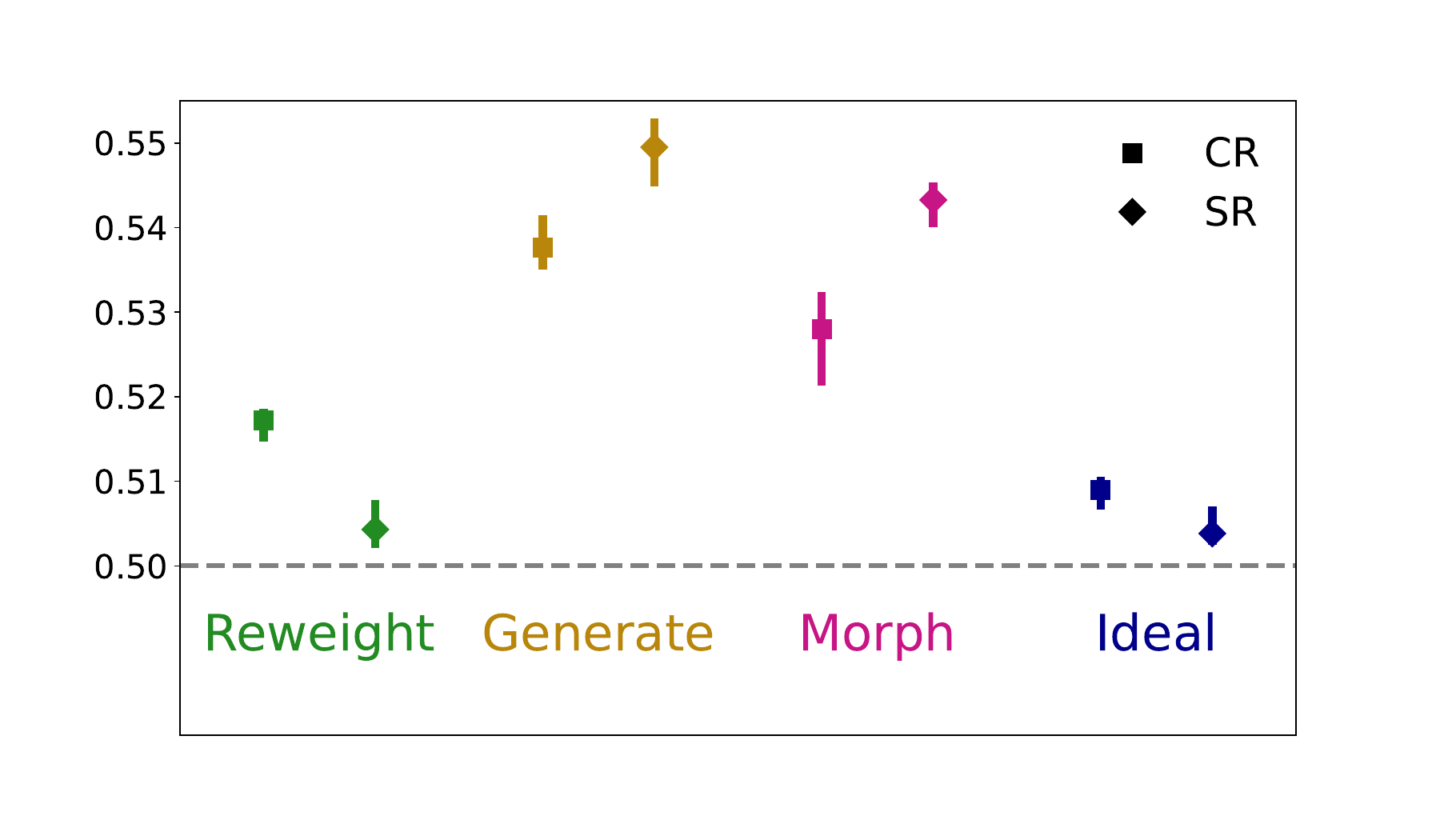}
    \caption{\label{fig:aucs_physics} ROC AUC spread in the CR and SR. The horizontal line is at 0.5, marking true random. }
  \end{subfigure}
  \begin{subfigure}[t]{0.495\textwidth}
    \centering
    \includegraphics[width=\linewidth]{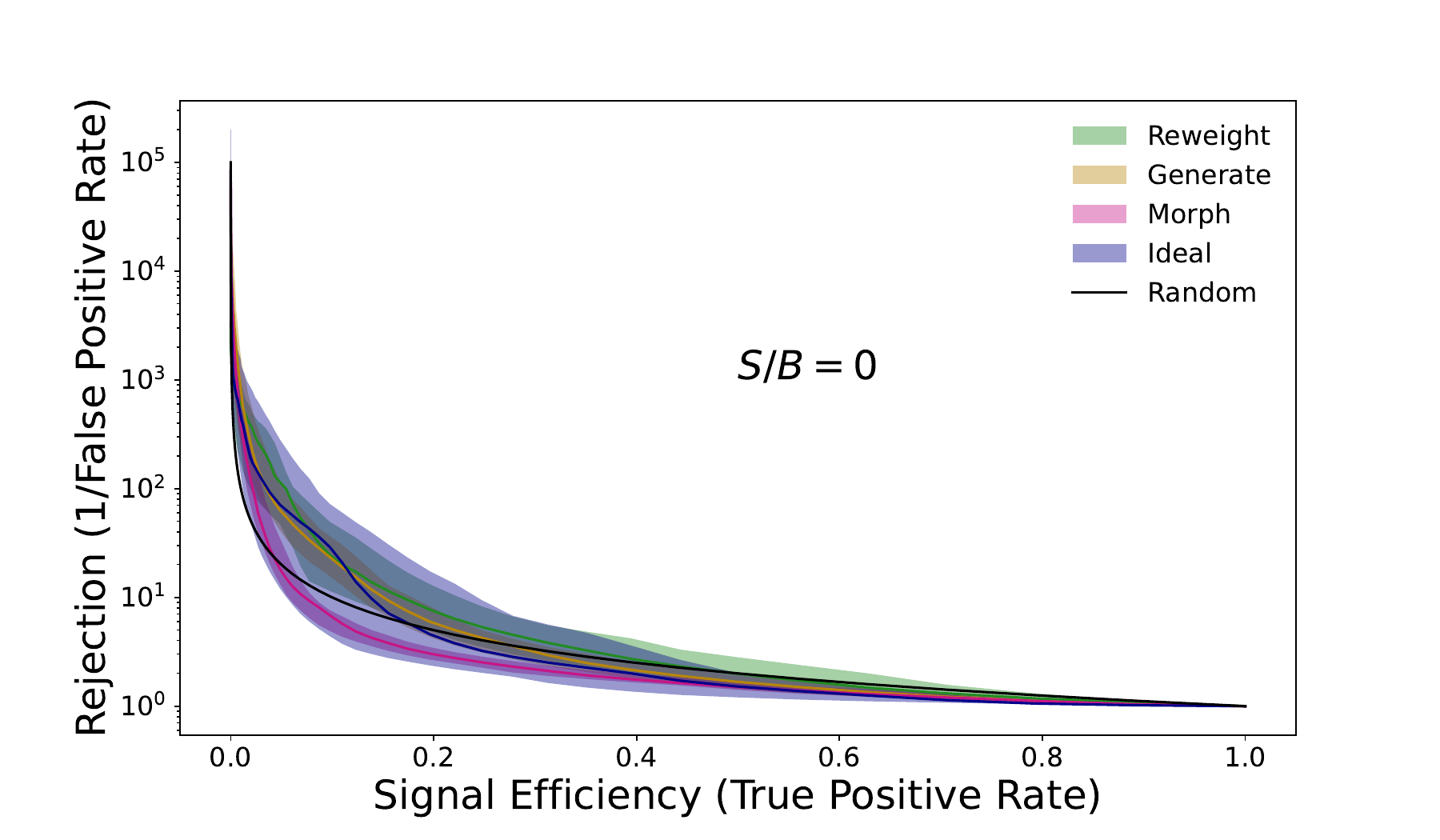}
    \caption{\label{fig:rej_0}Rejection in the SR.}
  \end{subfigure}
\caption{\label{fig:closure_plots}  Results for a classifier tasked with discriminating extrapolated background samples from true background. The ``Ideal'' corresponds to an idealized classifier, which has been trained to discriminate two statistically identical sets of background, and thus represents a realistic random classifier. Errorbars and uncertainty bands show a 68-percentile spread across 20 independent classifier runs. }
\end{figure}

In \Fig{fig:aucs_physics}, we show a more rigorous test by training a binary classifier to discriminate between the constructed samples and background-only data.  We show the spread of ROC AUCs (receiver operating characteristic area-under-curves) across $10$ binary classifier runs; for a classifier trained on two identical datasets, the ROC AUC is expected to be 0.5 for a perfect classifier, in the limit of infinite statistics and training time. Again, we see that there is generally excellent reconstruction in the CR and slightly worse reconstruction in the SR.  In \Fig{fig:rej_0}, we show the classifier rejection ($1 / \textrm{false positive rate}$) curves for the discrimination task in the SR. For all of the methods, the rejection found in the zero-signal case is compatible (to within errorbars) with that of a random classifier.

As stated in Sec. \ref{sec:NN_specifications}, architectures and hyperparameters for the extrapolation networks were selected so as to make the spread of ROC AUCs in the CR consistent with random. There is a (seemingly unavoidable) systematic error associated with the extrapolation, such that Reweight, Morph, and Generate methods show a drop in similarity to true background in the SR. We believe that most of this drop can be attributed to the more general challenge of extrapolation for neural networks.

\subsubsection{Signal detection tests}

In this section, we carry out an anomaly detection test for the dark QCD physics model. The test is done by generating a signal-background discriminator using a weekly-supervised binary classifier following the \textsc{CWoLa} procedure, as explained in Sec.~\ref{sec:methodology}. The classifier is trained to discriminate the background reference against data that consists of SR background with a known amount of signal injection. For the simulation, we scan over seven signal injections of signal ($S$) over background ($B$) of $S/B \in \{0, 3.1, 6.5, 9.3, 12, 15, 18\}\times 10^{-3}$, corresponding to significances $S/\sqrt{B} \in \{0, 0.84, 1.75, 2.52, 3.32, 4.07, 4.99$\}, respectively\footnote{Note that with our chosen context region bounds, there is a small amount of signal contamination in the CR. For the largest signal injection, the contamination is 4.3$\times 10^{-3}$\%, from 384 events.}. In this section, we provide a number of metrics derived from this binary classifier.

We focus on metrics derived from the classifier SIC curves for the discrimination task in the SR. The SIC, defined as the $\textrm{true positive rate} / \sqrt{\textrm{{false positive rate}}}$, gauges the factor by which a signal's significance would improve by making a cut on the classifier score.  We expect SIC $\gg 1$ for a well-performing classifier, while in the zero-signal case, the classifier would perform poorly. 

\begin{figure}
 \centering
  \begin{subfigure}[t]{0.495\textwidth}
    \includegraphics[width=\linewidth]{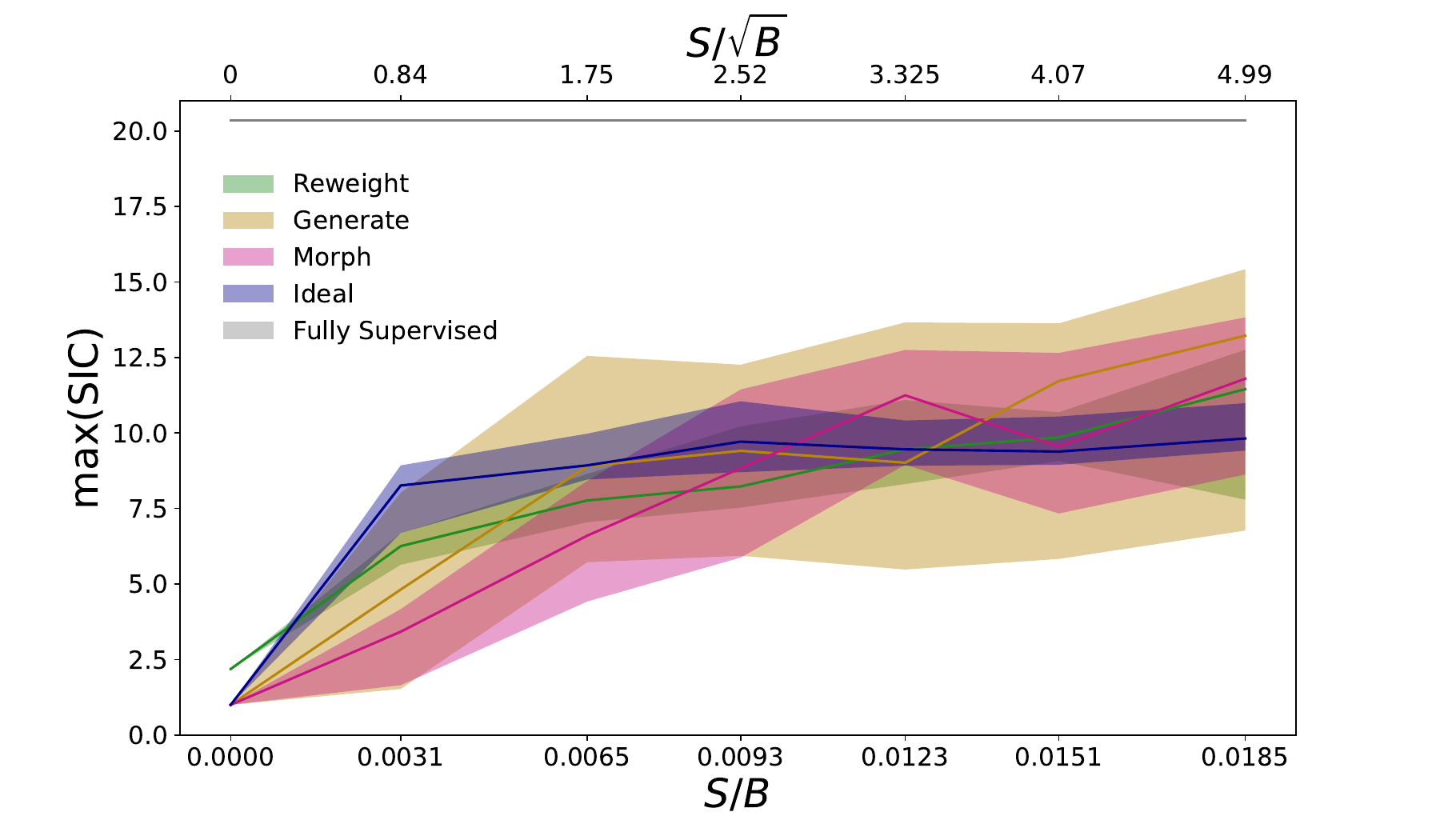}
    \caption{\label{fig:maxsic_reg_mc_4TeV} Maximum of the Significance Improvement Characteristic.}
  \end{subfigure}
  \begin{subfigure}[t]{0.495\textwidth}
    \centering
    \includegraphics[width=\linewidth]{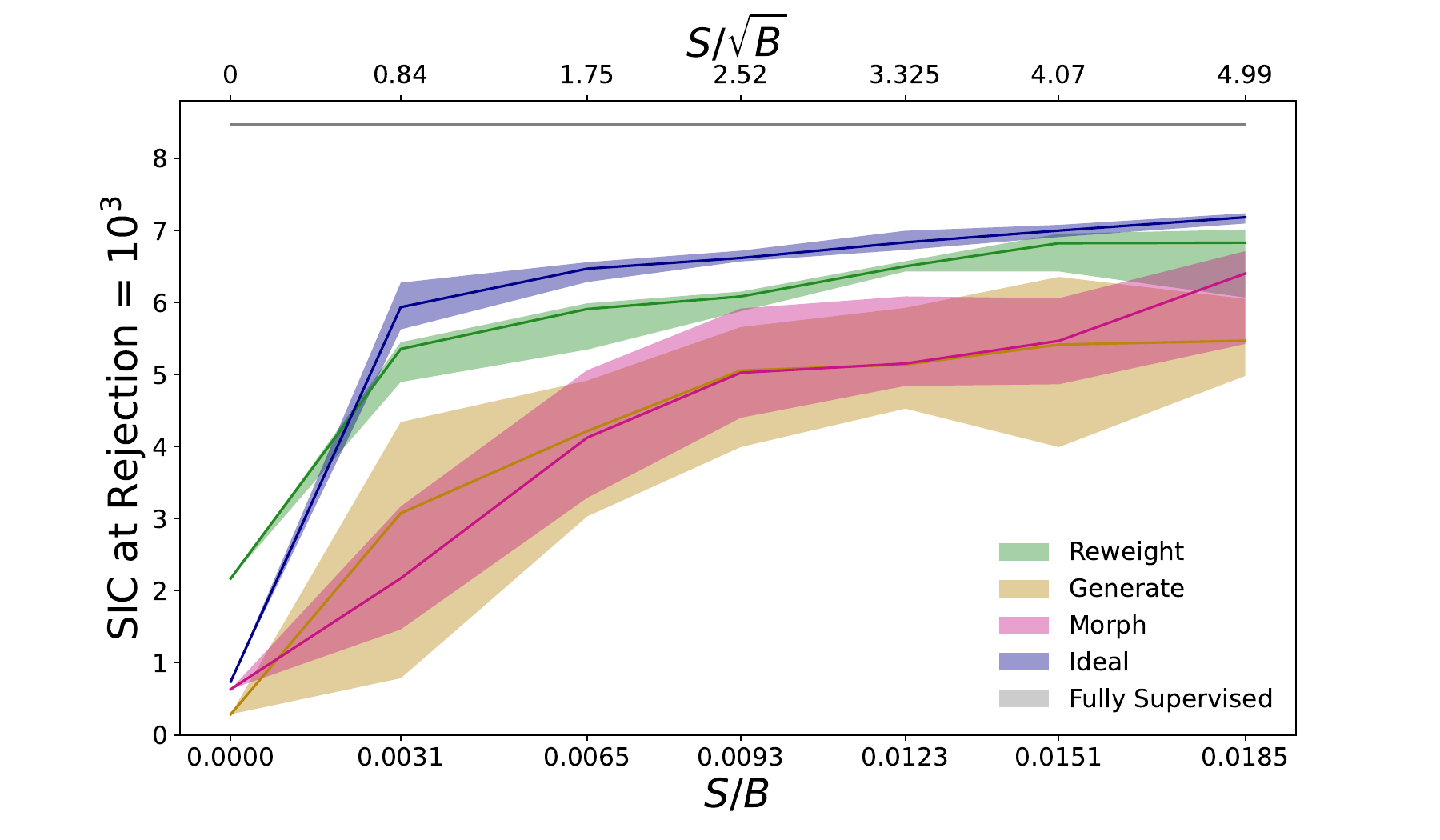}
    \caption{\label{fig:sic_at_rej_1000_reg_mc_4TeV} Significance Improvement Characteristic evaluated at a rejection of $10^3$.}
  \end{subfigure}
\caption{\label{fig:scan_plots_reg_mc_4TeV} Metrics for an anomaly detection classifier tasked with discriminating extrapolated background samples from data in the SR. Uncertainty bands represent the 68-percentile spread over 10 different signal injections, all of which have been score-averaged over 20 independent classifier runs. }
\end{figure}

In \Fig{fig:maxsic_reg_mc_4TeV}, we show the maximum SICs across all signal efficiencies, corresponding to the largest achievable significance improvements. In \Fig{fig:sic_at_rej_1000_reg_mc_4TeV}, we show SICs evaluated at a fixed rejection (given by $1 / \textrm{false positive rate}$) of $10^3$. The results are encouraging, demonstrating comparable performance to the idealized anomaly detector for all three methods. The Reweight method enhances a signal significance from below $1 (\sim 0.84) \sigma$ to the $5 \sigma$ ``discovery'' threshold. The Generate and Morph methods perform less optimally with respect to the metric of SIC at a rejection of $10^3$, but there is good evidence that the Generate and Morph methods could enhance a signal significance originally between $1$--$2\sigma$ to the discovery threshold. At zero signal injection, the higher SIC values for Reight is likely due to randomness, since there is nothing for the network to learn. What we see is that the mismodeling happens to be in a signal-like direction. It would be interesting for future studies to explore how these findings change for different signal and background models\footnote{See Appendix~\ref{appendix:a}, where we show how the detection task changes when using a different set of signal parameters.}.

\section{Conclusions}
\label{sec:conclusions}

In this paper, we have extended background interpolation methods to the extrapolation case to enable weakly supervised, non-resonant anomaly detection (AD).  Our strategy is based on the construction of a realistic set of background events through extrapolation from a background enriched control region (CR) into a signal region (SR). We proposed three extrapolation approaches based on reweighting, generation, and morphing which parallel the SALAD, CATHODE, and FETA methods, respectively.  Studies in the resonant case have shown complementarity between these methods~\cite{Golling:2023yjq} which may be especially important in the extrapolation case.

The proposed methods are tested on a realistic BSM signal consisting of dark QCD jets. We found that in the zero-signal case, there was good closure between the reconstructed samples and the background data, showing that our background templates were well-modeled and unbiased. The agreement between predicted background and the true background in CR implies our methods are optimized for interpolation, and deviation in the SR might come from the intrinsic difficulty of extrapolation. Finally, we have also shown the performances of all the methods in a realistic AD task, demonstrating their abilities to elevate non-significant signal injection to the threshold of possible detection.  While we only showed the performance for a single physics signal model, we stress that the protocol was optimized on only the background, and the sensitivity should extend to a broad class of new physics scenarios.

Our study, while encouraging, brings to light some of the many challenges associated with non-resonant AD. Future work is necessary in order to develop more robust non-resonant AD. Possible avenues may involve using networks or data preprocessing techniques that are especially well-suited for extrapolation. It is also worth testing the learning of $w(m)$ function given multiple different simulation sets that are qualitatively different from the data, in order to verify the effectiveness of the Reweight method\footnote{See Appendix~\ref{appendix:b} for an example of a modified simulation set.}.
 
With background extrapolation methods, the range of possible signals that could be detected using an overdensity analysis would be greatly expanded. Not only are we able to explore a larger parameter phase space of known resonance signals that was previously not reconstructable, but we are also able to access non-resonant off-shell effects from heavy particles.  It is likely that no one method will achieve the best sensitivity to all scenarios, so a set of techniques are required to achieve broad coverage.

\section*{Data and code availability}

The physics datasets and parameter cards used for all simulations have been made available on Zenodo at \url{https://zenodo.org/records/10154213}.  The analysis code is available at \url{https://github.com/hep-lbdl/non-resonant-AD-extrapolation}.

\section*{Acknowledgments}

We thank Timothy Cohen, Sascha Diefenbacher, Laura Jeanty, Simon Knapen, and Christiane Scherb for the useful discussions.  In particular, we thank Christiane Scherb for the help on Hidden Valley simulations, and Laura Jeanty for providing detailed feedback on the manuscript.  KB would like to thank Javier Montejo Berlingen for providing training and inspirations prior to this project.  KB is supported by the U.S. Department of Energy, Office of Science, Office of High Energy Physics program under Award Number DE-SC0020244.  KB's visit to LBNL for this collaboration is supported by the US ATLAS Center program.  BN and RM are supported by the U.S. Department of Energy (DOE), Office of Science under contract DE-AC02-05CH11231 and Grant No. 63038 from the John Templeton Foundation.  RM is additionally supported by Grant No. DGE 2146752 from the National Science Foundation Graduate Research Fellowship Program.
This research used resources of the National Energy Research Scientific Computing Center, a DOE Office of Science User Facility supported by the Office of Science of the U.S. Department of Energy under Contract No. DE-AC02-05CH11231 using NERSC award HEP-ERCAP0021099.  

\appendix

\section{Comparing different signal parameters}
\label{appendix:a}

In this section, we compare the performance of three proposed non-resonant AD methods (Reweight, Generate, and Morph) on different signal parameters of the dark QCD model. We choose the following two sets of parameters, both with $r_{\rm inv}=1/3$ in addition to the $4$ TeV $Z'$ signal used in Sec.~\ref{sec:physics_results}:

\begin{itemize}
    \item $m_{Z'}=2$ TeV, $m_{\pi_{\rm D}} = m_{\rho_{\rm D}} = \Lambda_\text{D} = 200$ GeV, and $m_{q_{\rm D}} = 100$ GeV.
    \item $m_{Z'}=3$ TeV, $m_{\pi_{\rm D}} = m_{\rho_{\rm D}} = \Lambda_\text{D} = 300$ GeV, and $m_{q_{\rm D}} = 150$ GeV.
\end{itemize}

These two sets of parameters, denoted as the $2$ TeV and $3$ TeV signal, respectively, are chosen to have different $m_{jj}$ and $H_{\rm T}$ distributions from the $4$ TeV signal, with more signal located towards the bulk of the background. The choice of $m_{\pi_{\rm D}}$, $ m_{\rho_{\rm D}}$, $\Lambda_\text{D}$ and $m_{q_{\rm D}}$ values gives a similar two-pronged and three-pronged jet substructures as the $4$ TeV signal in Sec.~\ref{sec:physics_results}. 

\begin{figure}[ht]
 \centering
  \begin{subfigure}[t]{.6\textwidth}
    \includegraphics[width=\textwidth]{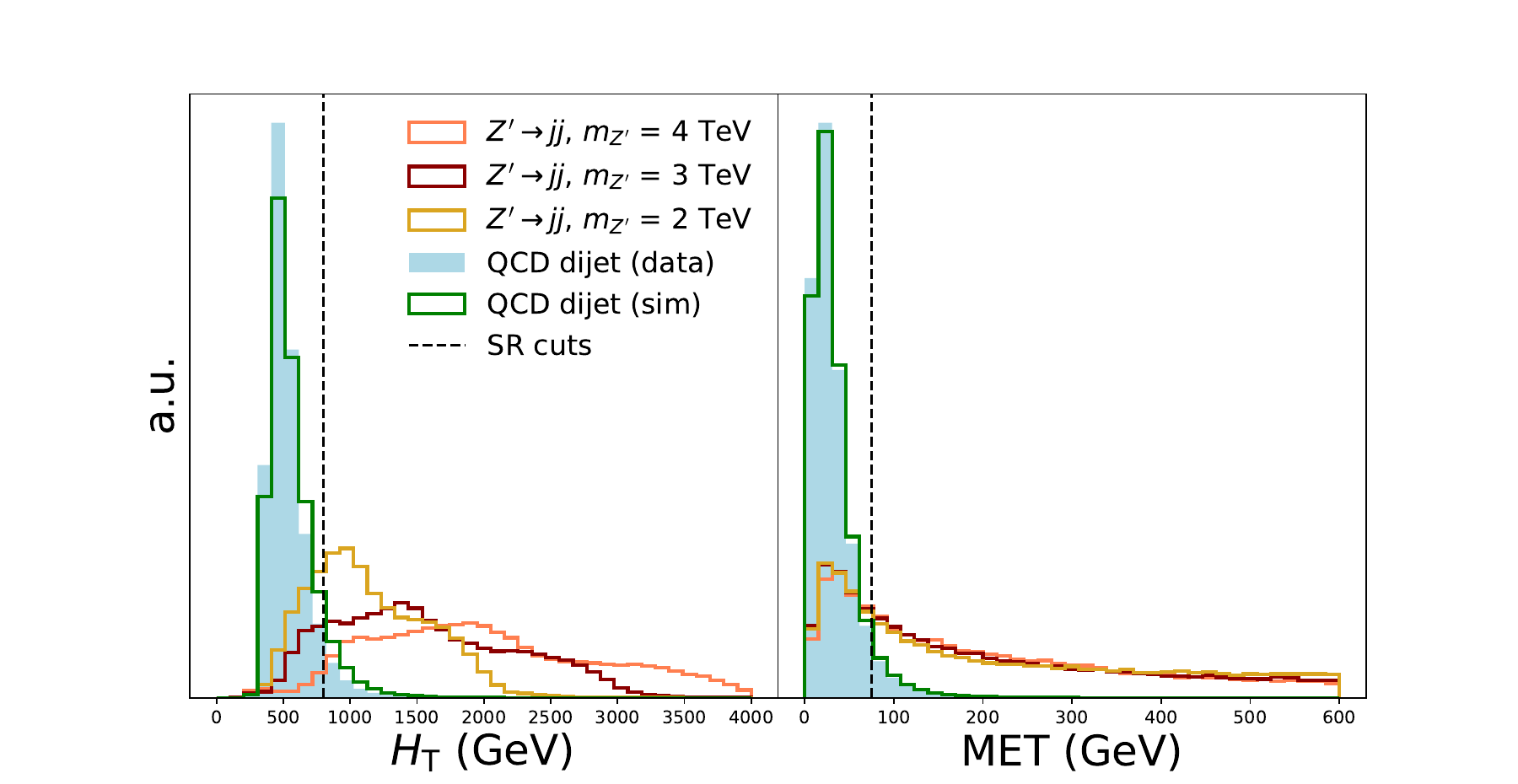}
    \caption{\label{fig:sig_vs_bkg_cont_allsig}Two-dimensional context variables}
  \end{subfigure}
  \\
  \hfill
  \begin{subfigure}[t]{1\textwidth}
    \centering
    \includegraphics[width=\textwidth]{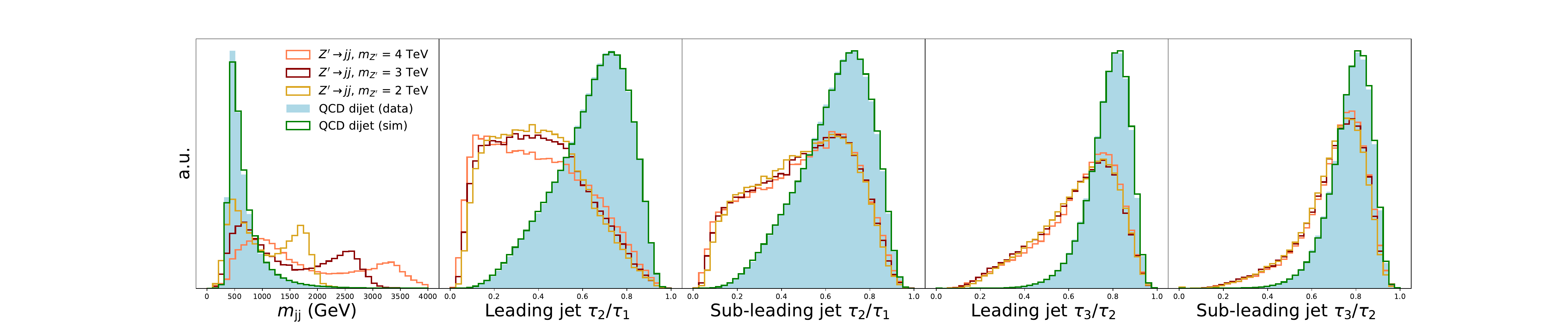}
    \caption{\label{fig:sig_vs_bkg_feat_allsig}Five-dimensional feature variables}
  \end{subfigure}
\caption{\label{fig:physics_datasets_allsig} Histograms of the observables used in the Hidden Valley non-resonant anomaly detection task, for background (``data'' and ``simulation'') and signal events). 
}
\end{figure}

Fig.~\ref{fig:physics_datasets_allsig} shows the distributions of context and feature variables of three sets of signal parameters used in this study. We apply the same SR definition with $H_{\rm T}>800$ GeV and MET $> 75$ GeV. Note that the distributions of the context variable $H_T$ and the feature variables $m_{jj}$ for the 2 TeV signal look much more similar to those of background, so we expect the discrimination task to be harder for lower signal masses.

\begin{figure}
 \centering
  \begin{subfigure}[t]{0.495\textwidth}
    \includegraphics[width=\linewidth]{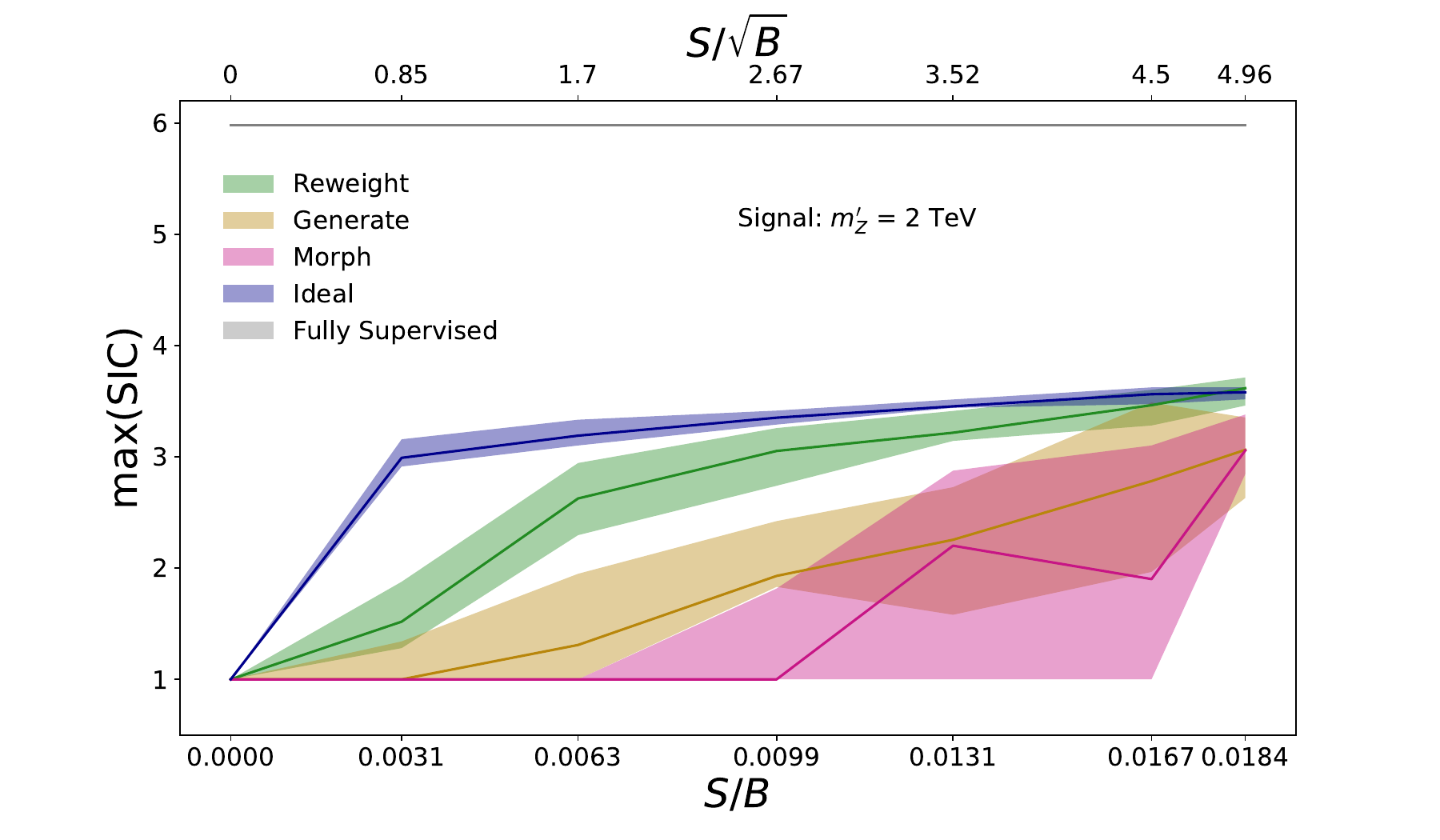}
    \caption{\label{fig:maxsic_reg_mc_2TeV} Maximum of the Significance Improvement Characteristic.}
  \end{subfigure}
  \begin{subfigure}[t]{0.495\textwidth}
    \centering
    \includegraphics[width=\linewidth]{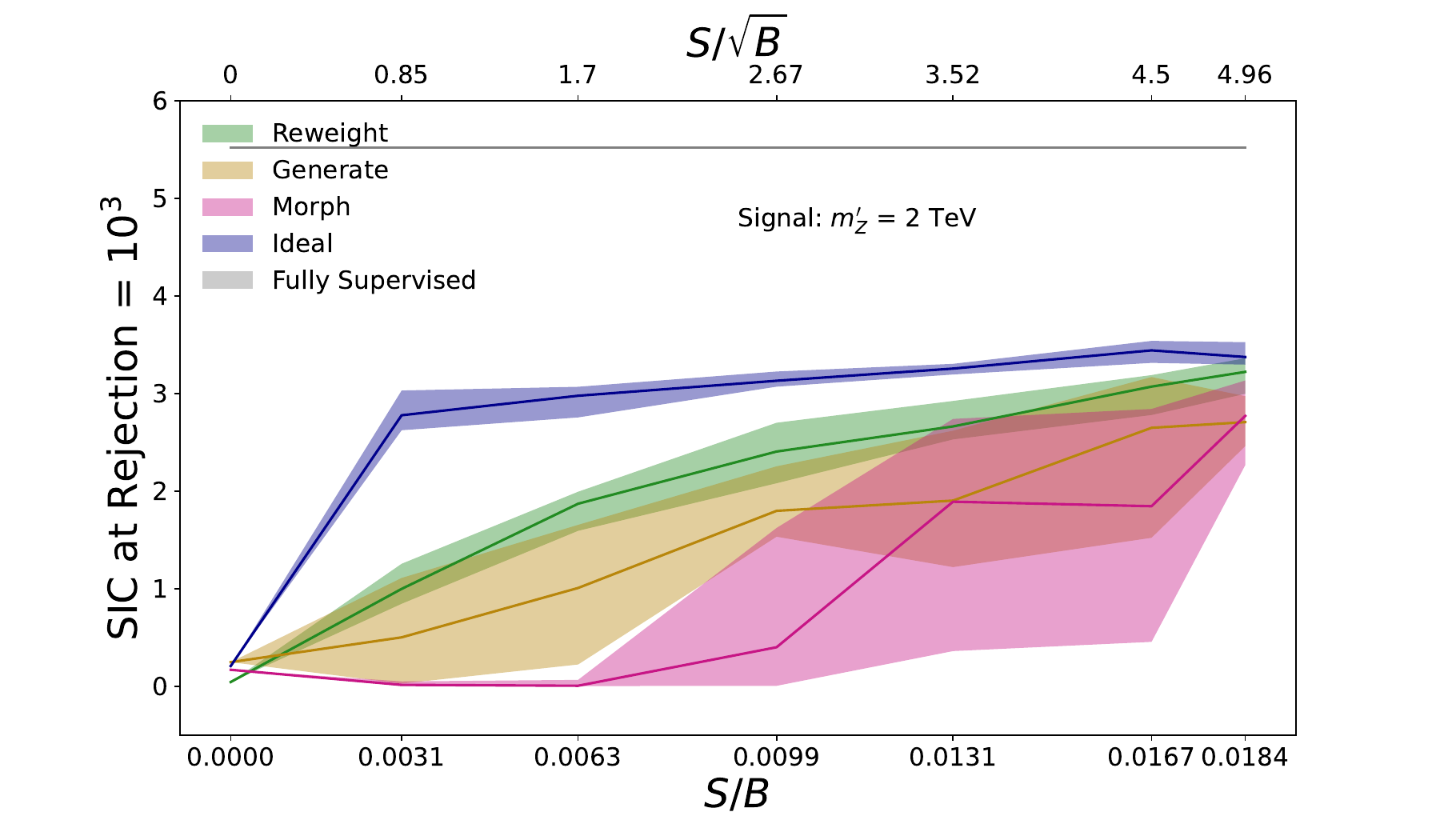}
    \caption{\label{fig:sic_at_rej_1000_reg_mc_2TeV} Significance Improvement Characteristic evaluated at a rejection of $10^3$.}
  \end{subfigure}
\caption{\label{fig:scan_plots_reg_mc_2TeV} Metrics for an anomaly detection classifier tasked with discriminating extrapolated background samples from data in the SR. The background is QCD dijet events. The signal is a $2$ TeV $Z'$ producing two dark QCD jets with $2 m_{q_{\rm D}}= m_{\pi_{\rm D}} = m_{\rho_{\rm D}} = \Lambda_\text{D} = 200$ GeV. Uncertainty bands represent the 68-percentile spread over 10 different signal injections, all of which have been score-averaged over 20 independent classifier runs. }
\end{figure}

\begin{figure}
 \centering
  \begin{subfigure}[t]{0.495\textwidth}
    \includegraphics[width=\linewidth]{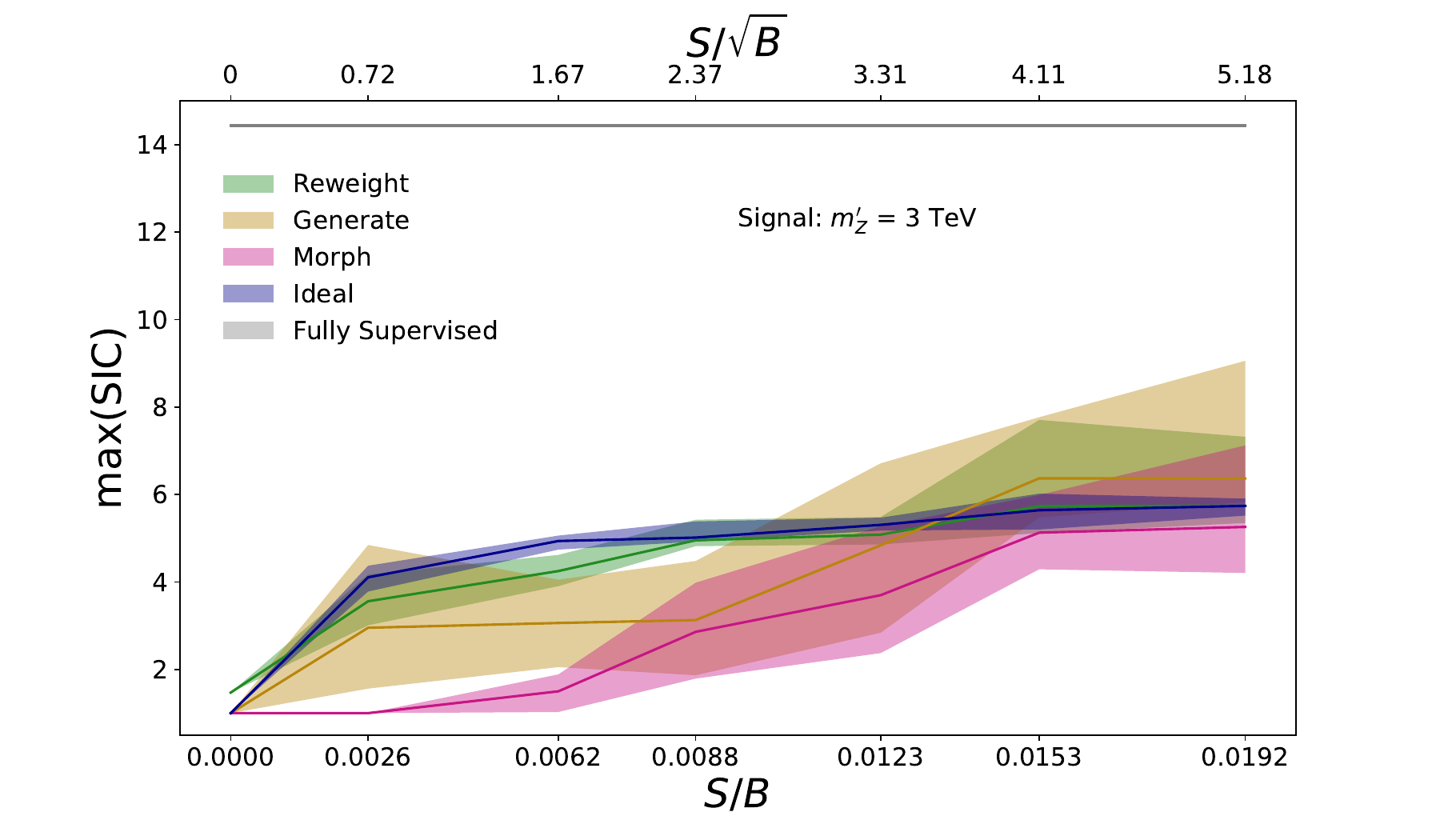}
    \caption{\label{fig:maxsic_reg_mc_3TeV} Maximum of the Significance Improvement Characteristic.}
  \end{subfigure}
  \begin{subfigure}[t]{0.495\textwidth}
    \centering
    \includegraphics[width=\linewidth]{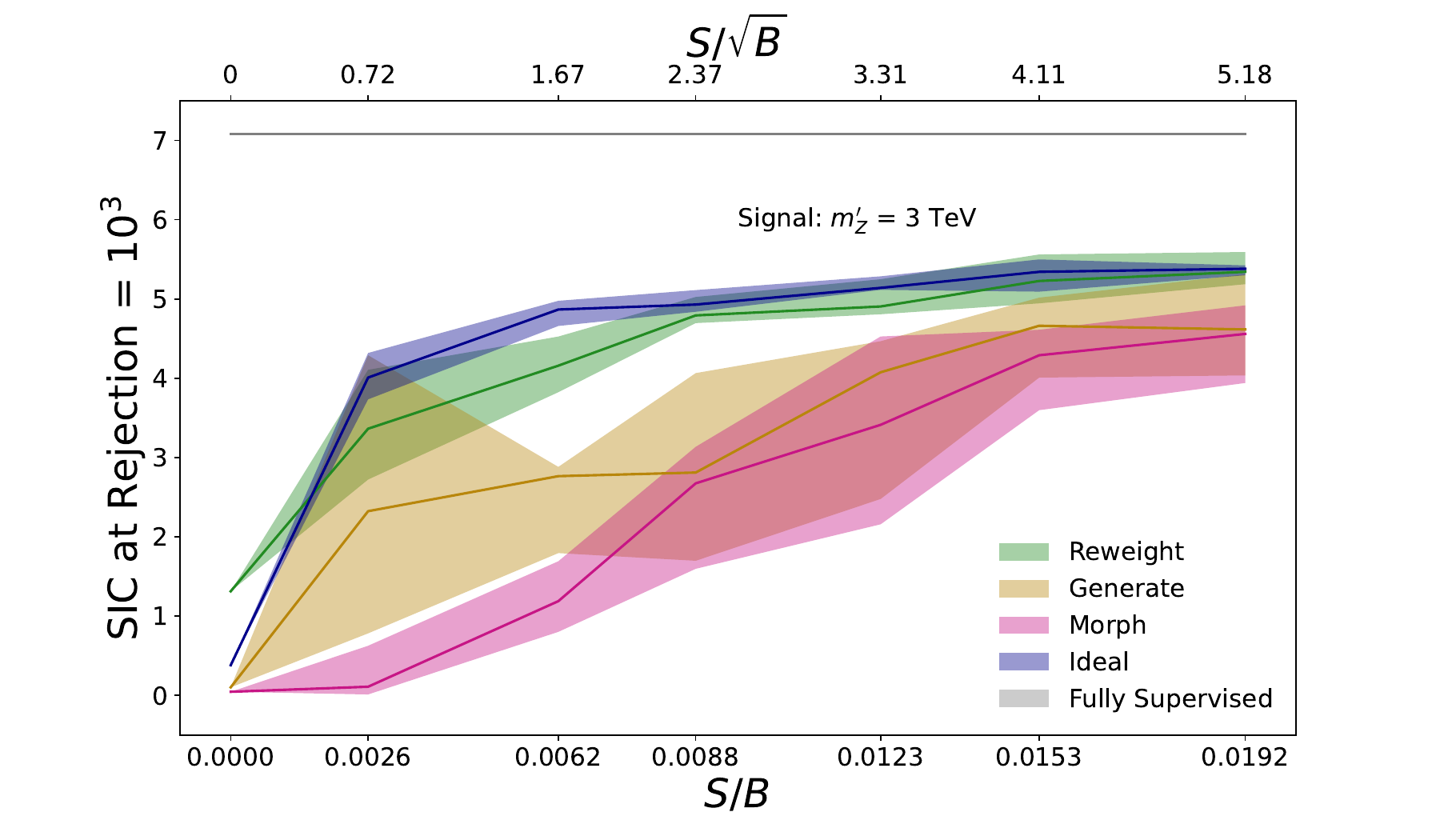}
    \caption{\label{fig:sic_at_rej_1000_reg_mc_3TeV} Significance Improvement Characteristic evaluated at a rejection of $10^3$.}
  \end{subfigure}
\caption{\label{fig:scan_plots_reg_mc_3TeV} Metrics for an anomaly detection classifier tasked with discriminating extrapolated background samples from data in the SR. The background is QCD dijet events. The signal is a $3$ TeV $Z'$ producing two dark QCD jets with $2 m_{q_{\rm D}}= m_{\pi_{\rm D}} = m_{\rho_{\rm D}} = \Lambda_\text{D} = 300$ GeV. Uncertainty bands represent the 68-percentile spread over 10 different signal injections, all of which have been score-averaged over 20 independent classifier runs. }
\end{figure}

In Figs.~\ref{fig:scan_plots_reg_mc_2TeV} and~\ref{fig:scan_plots_reg_mc_3TeV}, we show the results for an anomaly detection classifier working on a range of signal detections for the 2 TeV and 3 TeV signals, respectively. (These plots are analogous to \Fig{fig:scan_plots_reg_mc_4TeV} for the 4 TeV signal.) As expected, the performances of all AD methods show low sensitivity to the $2$ TeV signal, with Reweight achieving a similar significance improvement as the idealized classifier. Performance is better across the board for the 3 TeV signal, although the qualitative features are the same as for the 2 TeV signal.

For both 2 TeV and 3 TeV signals, the gap in performance between the idealized classifier and the fully supervised classifier is much larger than it is for the 4 TeV signal. This difference might be ascribed to the fact that the lower-mass signals have a greater resemblance to the background process in variables such as $H_T$ and $m_{jj}$. Additionally, the amount of signal contamination in the CR is higher for the lower mass $Z'$ signals: for the 2 TeV $Z'$, the highest signal injection has a CR contamination of $0.01\%$ (1257 events), and for the 3 TeV $Z'$, the highest contamination is $0.005\%$ (551 events). A significant contamination could bias the extrapolated background distributions.

However, there is still sufficient evidence that the Reweight method could reliably enhance the signal significance from 1$\sigma$ to the discovery threshold for all considered signal parameters. In contrast, the Generate and Morph methods appear to be more sensitive to the signal model, while still effective at picking up on moderate signal significances originating at $\sim 3\sigma$. It is worth exploring in which scenario the Reweight method becomes less effective than Generate and Morph in the future.

\section{Modifying the simulation}
\label{appendix:b}

In the main text of this work, the distributions of simulated background are similar to the data distributions in most variables. This feature could decrease the difficulty of the reweighting procedure in all three methods. In this section, we investigate the effects of using a ``morphed'' simulation set that might better represent the expected discrepancy between simulation and actual collider data. 

We introduce a hand-tuned modification to the simulation, both to the features and context, defined in Eq. \ref{eq:sim_mod}:

\begin{align}
\begin{split}
H_T:& \hspace{3mm} x \rightarrow x \\
\textrm{MET}:& \hspace{3mm} x \rightarrow x\left(1+\frac{x}{500}\right)\\
m_{jj}:& \hspace{3mm} x \rightarrow x\left(1+\frac{x}{6000}\right)\\
\tau_{i}:& \hspace{3mm} x \rightarrow x^{1.1}.\\
\label{eq:sim_mod}
\end{split}
\end{align}

\begin{figure}[ht]
 \centering
  \begin{subfigure}[t]{.6\textwidth}
    \includegraphics[width=\textwidth]{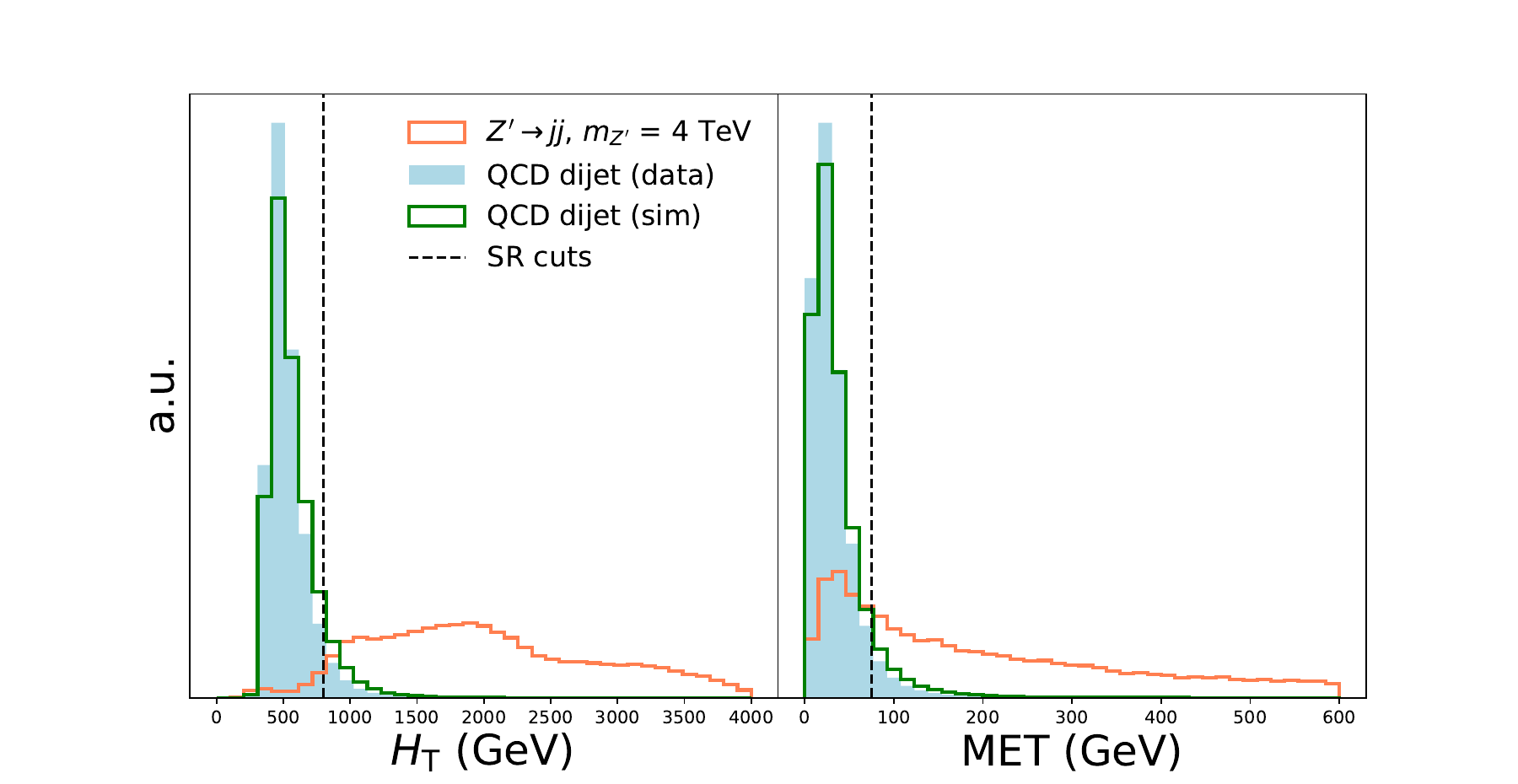}
    \caption{\label{fig:sig_vs_bkg_cont_morphmc}Two-dimensional context variables}
  \end{subfigure}
  \\
  \hfill
  \begin{subfigure}[t]{1\textwidth}
    \centering
    \includegraphics[width=\textwidth]{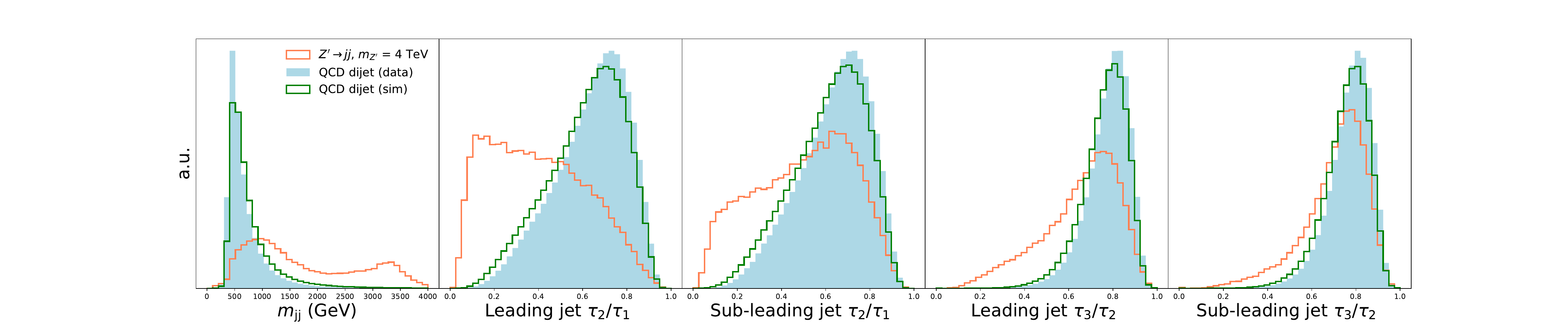}
    \caption{\label{fig:sig_vs_bkg_feat_morphmc}Five-dimensional feature variables}
  \end{subfigure}
\caption{\label{fig:physics_datasets_morphmc} Histograms of the observables used in the Hidden Valley non-resonant anomaly detection task, for background (``data'' and ``simulation'') and signal events). The QCD dijet simulation has been modified to appear different from QCD dijet data. 
}
\end{figure}

\noindent Through this modification, the distributions of simulation are slightly morphed to be different from data, as shown in \Fig{fig:physics_datasets_morphmc}.

\begin{figure}
 \centering
  \begin{subfigure}[t]{0.495\textwidth}
    \includegraphics[width=\linewidth]{plots/sic_at_rej_1000_reg_mc_4TeV}
    \caption{\label{fig:sic_at_rej_1000_reg_mc_4TeV_2} Metric for unmodified simulation (identical to \Fig{fig:sic_at_rej_1000_reg_mc_4TeV}).}
  \end{subfigure}
  \begin{subfigure}[t]{0.495\textwidth}
    \centering
    \includegraphics[width=\linewidth]{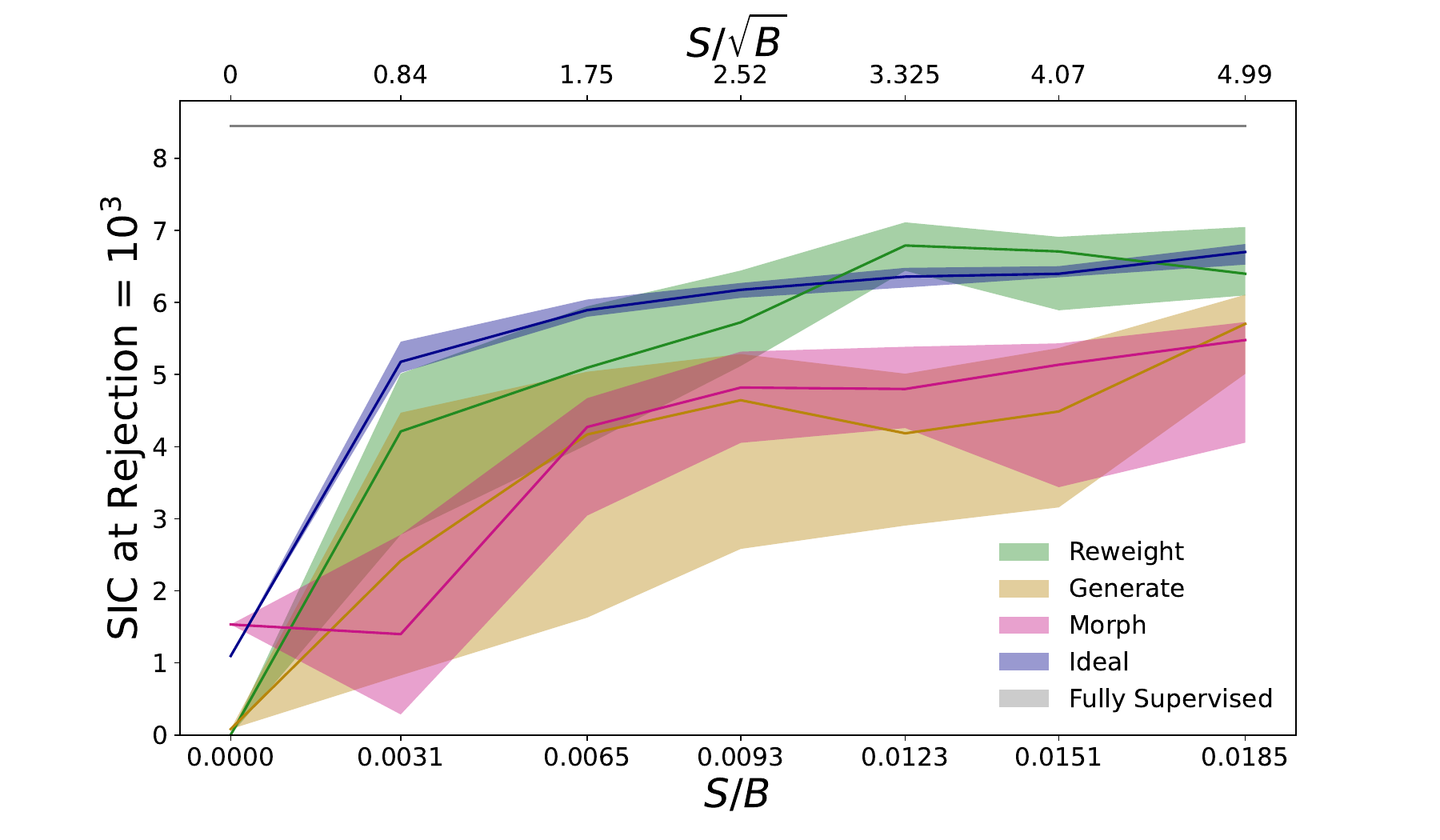}
    \caption{\label{fig:sic_at_rej_1000_mod_mc_4TeV} Metric for modified simulation}
  \end{subfigure}
\caption{\label{fig:scan_plots_mod_mc_4TeV} SIC at a rejection of $10^3$ for an anomaly detection classifier tasked with discriminating extrapolated background samples from data in the SR. Uncertainty bands represent the 68-percentile spread over 10 different signal injections, all of which have been score-averaged over 20 independent classifier runs. }
\end{figure}

In \Fig{fig:scan_plots_mod_mc_4TeV}, we compare the results for an anomaly detection classifier working on a range of signal detections for the modified simulation, using the same 4 TeV signal as was done in the main text. We make the comparison on the basis of the SIC at a rejection of $10^3$; in \Fig{fig:sic_at_rej_1000_reg_mc_4TeV_2}, we show the results for the unmodified simulation, and in \Fig{fig:sic_at_rej_1000_mod_mc_4TeV}, we show the results from modified simulation. The performances are compatible, both quantitatively and qualitatively. In particular, the signal injections at which each method reaches the detection threshold are almost equal. The main difference is that the uncertainty bands are wider for the morphed simulation. This is logical given that when the simulation is more different from the ``truth'' background, the networks must learn a less trivial function to map from signal to background.

\bibliography{HEPML,main}
\bibliographystyle{JHEP}

\end{document}